\documentstyle[epsfig]{elsart}
\date{2 September, 1999}
\begin{document}
\runauthor{Kuraev, Schiller, Serbo and Shaikhatdenov}
\begin{frontmatter}
\title{
\rightline{\small UL--NTZ 18/99}
Helicity amplitudes for the small--angle process
\boldmath $e^-e^{\pm}\to e^-\gamma\gamma + e^{\pm}$ \unboldmath
with both photons along one direction
and its cross channel \unboldmath }
\author[Dubna]{E.A.~Kuraev\thanksref{EAK}},
\author[Leipzig]{A.~Schiller\thanksref{AS}},
\author[Novosibirsk]{V.G.~Serbo\thanksref{VGS}},
\author[Dubna]{B.G.~Shaikhatdenov\thanksref{BGS}}
\thanks[EAK]{E--mail: kuraev@thsun1.jinr.ru}
\thanks[AS]{E--mail: schiller@tph204.physik.uni-leipzig.de}
\thanks[VGS]{E--mail: serbo@math.nsc.ru}
\thanks[BGS]{E--mail: sbg@thsun1.jinr.ru}

\address[Dubna]{Joint Institute for Nuclear Research, Dubna, 141980,
                Russia }
\address[Leipzig]{Institut f\"ur Theoretische Physik and NTZ,
                  Universit\"at Leipzig, D-04109 Leipzig, Germany,
                   }
\address[Novosibirsk]{Novosibirsk State University, Novosibirsk,
        630090, Russia}

\begin{abstract}
We study the small--angle double bremsstrahlung in  $e^-e^\pm$
scattering for a jet kinematics  where both photons  move
along the electron direction.  This region gives the main
contribution to the cross section.  We present analytic
expressions for all 64 amplitudes with arbitrary helicity states
of the initial and final leptons and the produced photons
convenient for analytic and numerical studies.
The accuracy of the obtained amplitudes is given omitting only
terms of the order of $ m^2/E_j^2$, $\theta_j^2$ and $\theta_j
m/E_j$.  The helicity amplitudes for the cross channel $\gamma
e^\pm \to \gamma e^+e^- + e^\pm$ are given. Several limits
for the helicity amplitudes of hard or soft final particles
are considered.

\end{abstract}


\end{frontmatter}

\section{Introduction}
Nowadays colliders and detectors are able to work with
polarized particles. Therefore
QED jet--like processes are required to be described
in full detail including the helicities of all particles.
Among other reactions, those inelastic processes
whose cross sections do not fall with energy are of special interest.
In particular, QED jets formed by two or three particles
in reactions of the type
\begin{eqnarray*}
  e^+ e^- \to e^+ \gamma + e^-\,, \quad e^+ e^- \to e^+ \gamma
  \gamma + e^-\,, \quad e^+ e^- \to e^+e^+ e^- + e^-
\end{eqnarray*}
provide a realistic model for hadronic jets.
The complete set of all QED jet--like tree processes up to order
$\alpha^4$ in the electromagnetic
coupling have been listed in our previous paper \cite{KSSS}.
For all of them except one we have obtained compact and simple
analytic expressions for all helicity amplitudes to a high
accuracy \cite{KSS85,KSS86,KSSS}.
The process of double bremsstrahlung along one direction and
its cross channel, discussed in the present paper, completes the
calculation of all helicity amplitudes to order $\alpha^4$.

\begin{figure}[!htb]
  \begin{center}
    \epsfig{file=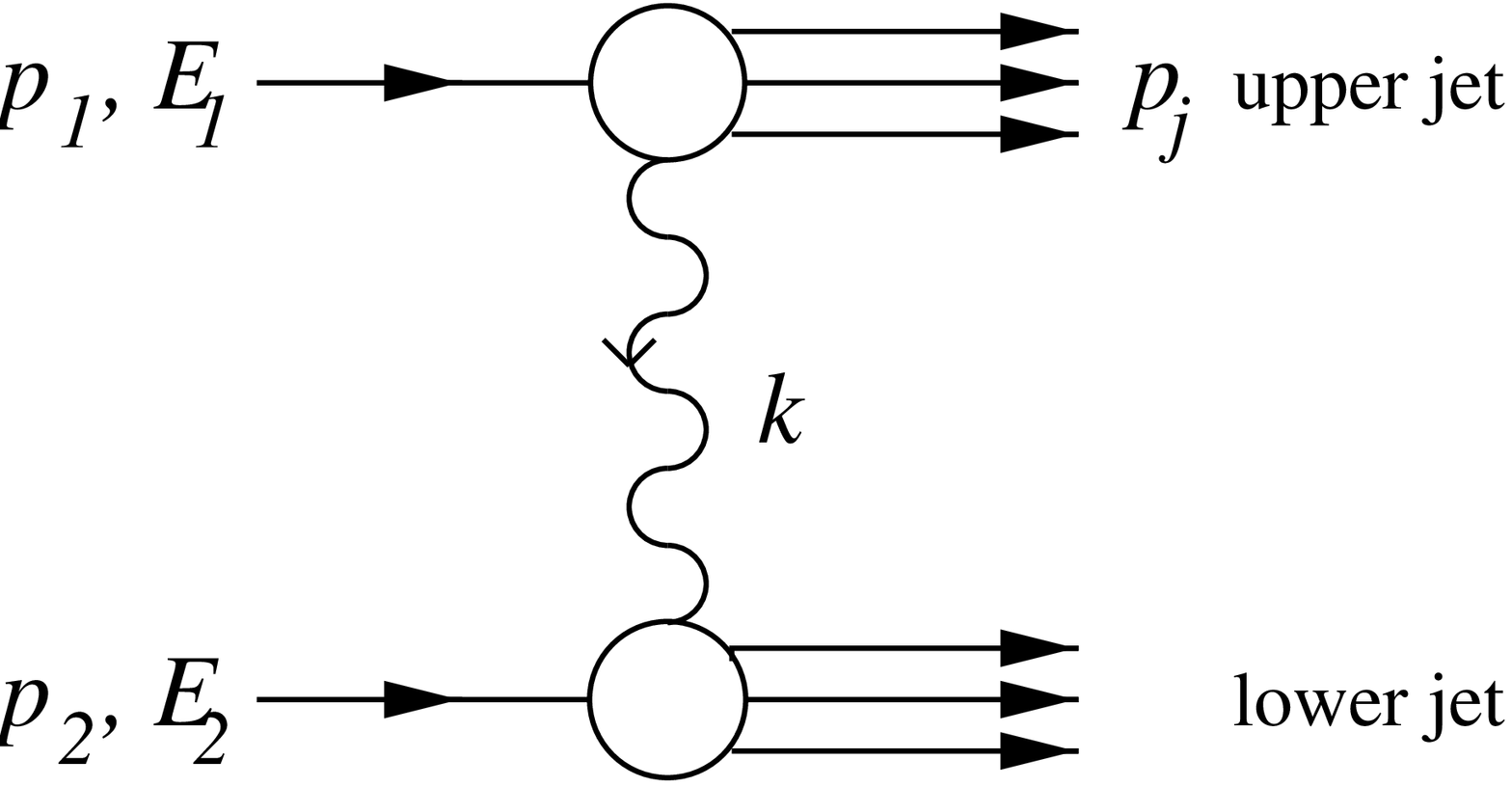,height=60mm,angle=270}
    \caption{ Generic block diagram  $e e \to $ two jets.}
    \label{fig:1}
  \end{center}
\end{figure}

Let us  define by a jet kinematics in  QED a high energy
reaction in which the outgoing leptons and photons are
produced within a small cone\footnote{The opening angle of the
cone is characterized by the maximal polar angles of the produced
particles.} relative to an axis given by their parental incoming
lepton or photon.  At high energies with the condition ($p_1$,
$p_2$ are the incoming 4-momenta, $m_j$ are the lepton masses)
\begin{eqnarray}
  s= 2 p_1 p_2 = 4 E_1 E_2 \gg m_j^2
  \label{mass}
\end{eqnarray}
the dominant contribution to the non-decreasing cross sections
is given by the region of scattering angles $\theta_j$ which is
much smaller than unity though could be of the order of the
typical emission angles $m_j / E_j$ or larger:
\begin{eqnarray}
  {m_j}/{E_j} \stackrel{<}{\sim} \theta_j  \ll 1 \ .
  \label{theta}
\end{eqnarray}
In this kinematic region all processes have the form of two--jet
processes with an exchange of a single virtual photon $\gamma^*$
 in the $t$--channel (see Fig.~\ref{fig:1}).

In the kinematic region (\ref{mass}), (\ref{theta}) it is possible
to obtain compact and relatively simple analytic expressions for
all helicity amplitudes of the jet-like QED processes to a high
accuracy omitting terms of the relative order of
\begin{eqnarray}
  \frac{m_j^2}{E_j^2} \,, \ \ \theta_j^2 \,, \ \
  \frac{m_j}{E_j} \,\theta_j
  \label{accuracy}
\end{eqnarray}
only.

The amplitude $M_{fi}$ corresponding to the diagram of
Fig.~\ref{fig:1} can be presented in the form
\begin{eqnarray} \label{9}
  M_{fi} = M_{\mathrm{up}}^{\mu} \,  \frac{g_{\mu \nu}} {k^2}  \,
  M_{\mathrm{down}}^\nu\,,
\end{eqnarray}
where $M_{\mathrm{up}}^\mu$ and $M_{\mathrm{down}}^\nu$
are the amplitudes of the upper and lower block of Fig.~\ref{fig:1},
respectively, and $(-g_{\mu \nu}/2)$ is the density matrix of the
virtual photon. The transition amplitude $M_{\mathrm{up}}$
describes the scattering of an incoming particle
(in our case a lepton) with a virtual photon of ``mass'' squared
$k^2$ and polarization 4--vector  $e=k_\perp/\sqrt{-k_\perp^2}$
($k_\perp$ is the 4--vector transverse to $p_1$ and $p_2$)
to some QED final state in the jet kinematics of Eqs.~(\ref{mass}),
(\ref{theta}) (similar for $M_{\mathrm{down}}$).

To introduce some notations for $e^-e^+$ collisions we use the
block diagram of Fig.~\ref{fig:1} as an example. We work in a
reference frame in which the initial particles with 4--momenta
$p_1$ and $p_2$ perform a head--on collision with energies $E_1$
and $E_2$ of the same order. The $z$--axis is chosen along the
momentum of the first initial particle, the azimuthal angles are
denoted by $\varphi_j$ (they are referred to one fixed $x$-axis).
Exploiting the freedom in choosing a general phase factor, we put
the initial azimuthal angle of the electron $\varphi_1$ and that of
the positron $\varphi_2$ equal to zero:
\begin{eqnarray}
  \varphi_1=\varphi_2=0 \,.
  \label{azim}
\end{eqnarray}
It is convenient to introduce ``the almost light--like'' 4--vectors
$p$ and $p'$
\begin{eqnarray}
  p &=& p_1 - \frac{m^2}{s} \, p_2\,, \quad
  p'=p_2 - \frac{m^2}{s}\, p_1\,, \quad p_1^2=p_2^2=m^2 \,,
\nonumber
\\
  p^2 & =& {p'}^2 = \frac{m^6} {s^2} \approx 0 \,, \ \
  s =  2p_1 p_2 = 2 p p' +  3 \frac{m^4}{s} \,.
  \label{light}
\end{eqnarray}

With accuracy (\ref{accuracy}) the matrix $g_{\mu\nu}$ from
Eq.~(\ref{9}) can be transformed to the form
\begin{eqnarray} \label{eq:2}
  g_{\mu\nu} \rightarrow
  \frac{2}{s} p'_{\mu}p_{\nu}\,,
\end{eqnarray}
(for details, see \cite{AB}, \S 4.8.4) which results in
the factorized representation for the amplitude
\begin{eqnarray}
  M_{fi}=\frac{s}{k^2} J_{\mathrm{up}} J_{\mathrm{down}} \,.
  \label{amplitude}
\end{eqnarray}
The vertex factors $J_{\mathrm{up,down}}$ are given by the block
amplitudes $M_{{\mathrm{up}}}^\mu$ and $M_{{\mathrm{down}}}^\nu$
\begin{eqnarray}
  J_{\mathrm{up}} = \frac{\sqrt{2}}{s} \,
  M_{\mathrm{up}}^\mu \, p'_\mu \,, \ \
  J_{\mathrm{down}} = \frac{\sqrt{2}}{s} \,
  M_{\mathrm{down}}^\nu \, p_\nu \,.
  \label{J1J2}
\end{eqnarray}
The quantities $J_{\mathrm{up}}$ and $J_{\mathrm{down}}$ can be
calculated in the limit $s \to \infty$ assuming that the energy
fractions and transverse momenta of the final particles are finite.

We study the double bremsstrahlung  in one direction for
small angle $e^-e^\pm$ scattering
\begin{eqnarray}
  e^-(p_1)+e^{\pm}(p_2)\to e^-(p_3)+\gamma(k_1)+\gamma(k_2)+e^
  {\pm}(p_4) \,.
  \label{DB}
\end{eqnarray}
For unpolarized particles this process has been discussed in
Refs.~\cite{BG} and \cite{KYF74} where the differential cross
sections and photon spectra were calculated in the approximation
of classical currents and equivalent photons.
In Refs.~\cite{KYF74,KF78} the matrix element squared for the
unpolarized case has been found to the accuracy (\ref{accuracy}).
The helicity amplitudes of the process (\ref{DB}) for
large angle scattering have been calculated in Refs.~\cite{KP,Behr}.
The recent interest to multiphoton emission is related to the exact
calculation of Bhabha scattering needed for LEP luminosity
calibration (see, for example, Refs.~\cite{CERN}).

\begin{figure}[!htb]
  \begin{minipage}{4.5cm}
    \begin{center}
       \epsfig{file=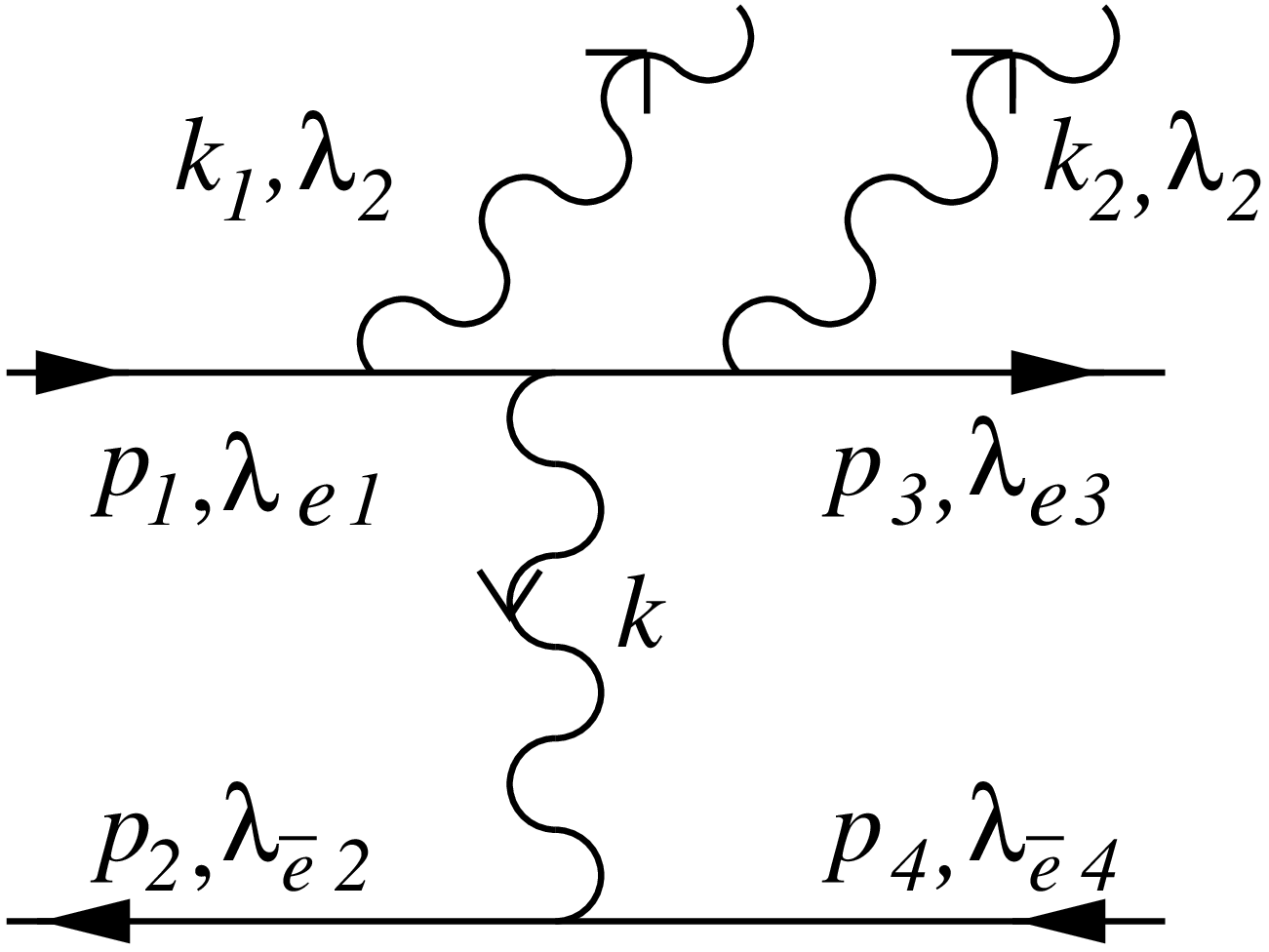,height=40mm,angle=270}
    \end{center}
  \end{minipage}
  \begin{minipage}{4.5cm}
    \begin{center}
       \epsfig{file=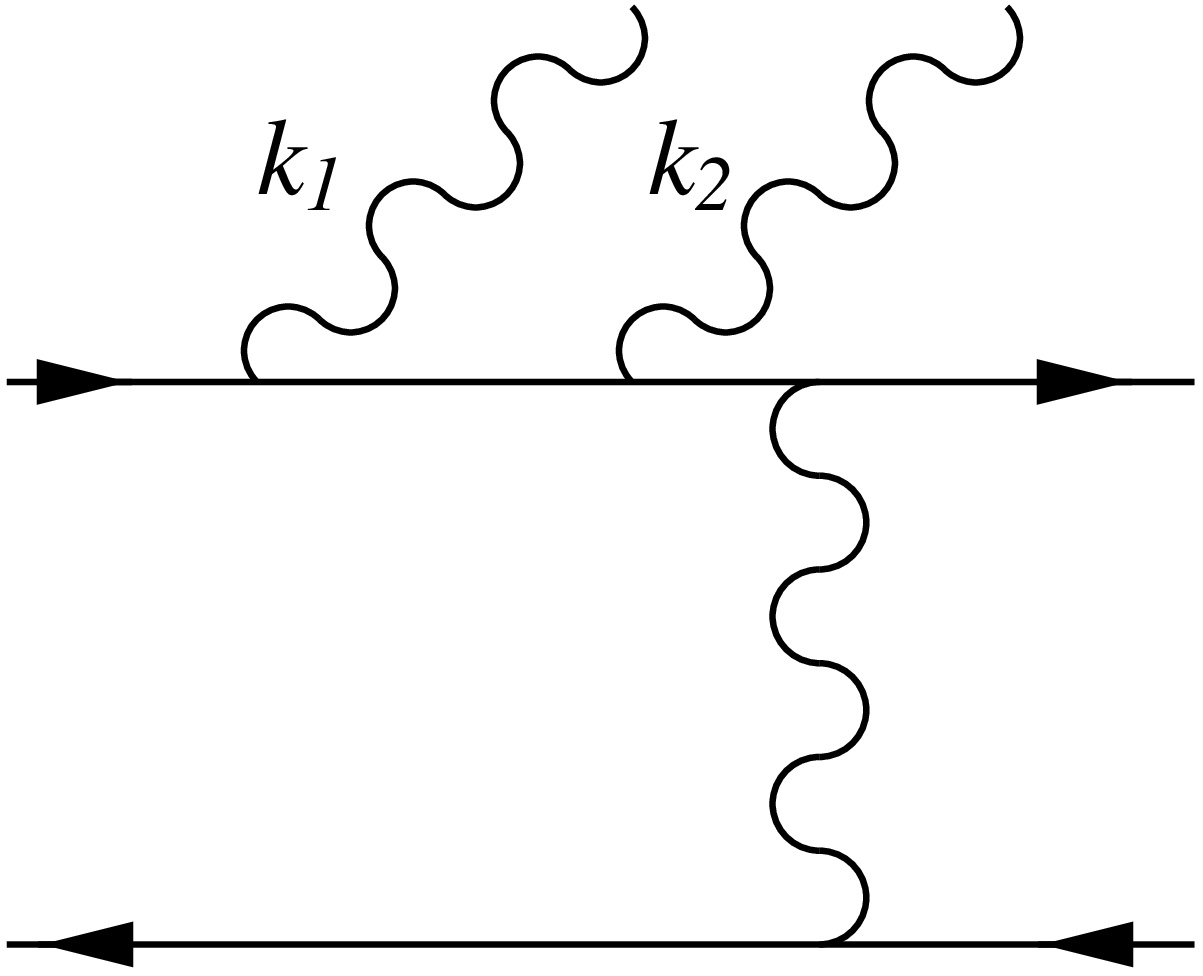,height=40mm,angle=270}
    \end{center}
  \end{minipage}
  \begin{minipage}{4.5cm}
    \begin{center}
       \epsfig{file=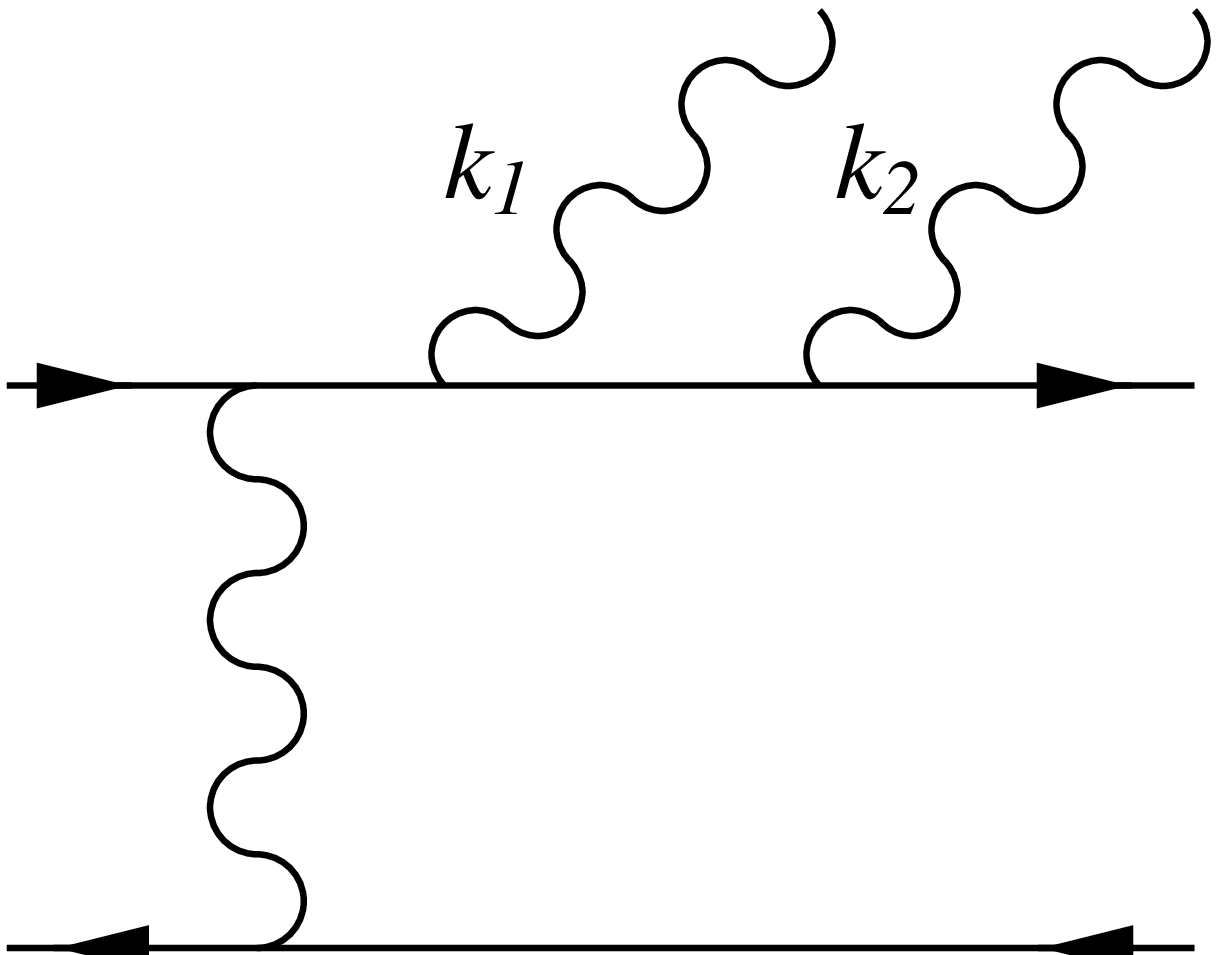,width=40mm,angle=0}
    \end{center}
  \end{minipage}
  \caption{Feynman diagrams for double bremsstrahlung in $e^- e^+$
           scattering, diagrams with $k_1\leftrightarrow k_2$ have
           to be added.}
  \label{fig:2}
\end{figure}

In Born approximation the process (\ref{DB}) (with photons not only
along one direction) is described by 32 Feynman diagrams. In the
considered kinematics (forward scattering at high energies) only
16 scattering--type diagrams are relevant since the contribution to
the cross section which does not decrease with the energy arises
from those Feynman diagrams with a virtual photon exchange
in the $t$--channel. Restricting to the case in which both photons
are emitted along the electron ($p_1$) direction, only six diagrams
are relevant. Three of them are shown in Fig.~\ref{fig:2}.

%%%%%%%%%%%%%%%%%%%%%%%%%%%%%%%%%%%%%%%%%%%%%%%%%%%%%%%%%%%%%%%%%%%%%%%%%%

\section{The double bremsstrahlung helicity amplitudes in the jet
kinematics}

\subsection{The Sudakov 4-vector decomposition and kinematic
invariants}

The amplitude for the double forward bremsstrahlung of the process
(\ref{DB}) can be written with accuracy (\ref{mass}), (\ref{theta})
in the factorized form (for definiteness the Bhabha
scattering is considered, see Fig.~\ref{fig:2}, where the momenta and
helicities of all particles are indicated)
\begin{eqnarray}
  &&M(e^-e^+\to e^-\gamma\gamma + e^+)
  =\frac{s}{k^2} J_{\mathrm{up}}({e^- \gamma^*\to e^- \gamma \gamma})
  J_{\mathrm{down}}( e^+ \gamma^* \to e^+) \,,
\\
  && k=p_4-p_2=p_1-p_3-k_1-k_2 \,.
  \nonumber
\end{eqnarray}
The vertex factor $J_{\mathrm{down}}$ has been calculated
earlier~\cite{KSSS}:
\begin{eqnarray}
  J_{\mathrm{down}} (e^+_{\lambda_{\bar{e}2}}
  \gamma^* \to e^+_{\lambda_{\bar{e}4}})=\sqrt{8 \pi \alpha}\,
  {\mathrm{e}}^{-{\mathrm{i}} \, \lambda_{\bar{e}4} \, \varphi_4}
  \,\delta_{\lambda_{\bar{e}2},\lambda_{\bar{e}4}} \,,
  \label{matelemdown}
\end{eqnarray}
The quantity $\lambda_{\bar{e}2}= \pm1/2$ ($\lambda_{\bar{e}4}$) denotes the
helicities of the initial (final) positron, $\varphi_4$ is the
azimuthal angle of the final $e^+$.

We have to calculate the upper vertex $J_{\mathrm{up}}$. The almost
light-like vectors  $p$ and $p'$ have components $ p=E_1(1,0,0,1)$,
$p'=E_2(1,0,0,-1)$. It is  convenient to use
the Sudakov decomposition of any 4--vector along $p$ and $p'$
and transverse to them. For the momenta of the photons and the
initial and final electron we have
\begin{eqnarray}
\label{sudakov}
  && k_i=\alpha_i p'+x_i p+k_{i\perp} \,,  \quad
      k_{i\perp}^2=-{\mathbf {k}}_i^2  \,,  \quad (i=1,2) \,,
\nonumber
\\
  && k=\alpha_k p'+\beta_k p+k_{\perp} \,, \quad
     k_\perp^2=-{\mathbf {k}}^2        \,,
\\
  && p_1= \frac{m^2}{s} p' +  p \,, \quad
     p_3= \alpha_3 p' + x_3 p + p_{3\perp} \,, \quad
     p_{3\perp}^2=-{\mathbf {p}}_3^2\,,
\nonumber
\end{eqnarray}
with the two--dimensional Euclidean vectors
${\mathbf{k}}$, ${\mathbf{k}}_i$ and
${\mathbf{p}}_3=-{\mathbf{k}}-{\mathbf{k}}_1- {\mathbf{k}}_2$
perpendicular to the beam axes, e.g.
\begin{eqnarray}
  {\mathbf k}=(k_x,k_y)\,.
  \label{2vector}
\end{eqnarray}
In the kinematics of a jet moving close to the direction of the
first electron the 2--vectors ${\mathbf {k}}_i$ and
${\mathbf {p}}_3$ have typical values of a few electron masses
whereas ${\mathbf {k}}$ might be much smaller. The quantities
$x_{1,2}$ and $x_3=1-x_1-x_2$ are the energy fractions
of the emitted photons and the scattered electron
\begin{eqnarray}
  x_i=\frac{k_{i0}}
  {E_1},\qquad x_3=\frac{p_{30}}{E_1}=\frac{E_3}{E_1} \,.
\end{eqnarray}
They are supposed to be of the order of unity while $\alpha_i$,
$\alpha_3$, $\alpha_k$ and $\beta_k$ are small and vanish as $1/s$.
Later on we often use complex circular components of a generic
2--vector transverse to the beam axes denoted by the corresponding
Greek characters, e.g. for the 2--vector (\ref{2vector})
\begin{eqnarray}
  \kappa= k_x + {\mathrm{i}} \, k_y \,, \quad
  \kappa^*= k_x - {\mathrm{i}} \, k_y \,.
  \label{circular}
\end{eqnarray}

The on--shell conditions for the incoming electron and the outgoing
electron and photons lead to
\begin{eqnarray}
  \label{alphai}
  &&s\alpha_i=\frac{{\mathbf {k}}^2_i}{x_i} \,, \quad
    s\alpha_3=\frac{m^2+ {\mathbf{p}}_3^2}{x_3}\,,
\\
  &&s(\alpha_k+\alpha_1+\alpha_2)=-\frac{1}{x_3}
    \left[ m^2(1-x_3)+{\mathbf p}_3^2 \right] \,.
  \nonumber
\end{eqnarray}
{}From the on--shell condition for the outgoing spectator (a positron
in Bhabha scattering) $p_4^2=(p_2+k)^2=m^2$ we find the expression
\begin{eqnarray}
  \qquad s\alpha_k\beta_k-{\mathbf {k}}^2+s\beta_k+m^2\alpha_k=0
  \label{sab}
\end{eqnarray}
and present the momentum transfer squared (the 4--momentum of
the virtual photon $k$) in the form
\begin{eqnarray}
  k^2=s\alpha_k\beta_k-{\mathbf {k}}^2=
      -\frac{1}{1+\alpha_k}\left({\mathbf{k}}^2
      +m^2\alpha_k^2\right) \,.
  \label{alpha}
\end{eqnarray}
Taking into account Eq.~(\ref{alpha}), we observe that the Sudakov
parameter $\alpha_k$ is related to the invariant mass squared of the
jet generated by the scattered electron and two accompanying photons
\begin{eqnarray} \label{invmass}
(p_3+k_1+k_2)^2=(p_1-k)^2=m^2-{\mathbf {k}}^2-s\alpha_k \,.
\end{eqnarray}
{}From Eqs.~(\ref{alphai}), (\ref{sab}) and (\ref{alpha}) explicit
expressions for $\alpha_k$ and $\beta_k$ can be derived.

After replacing all Sudakov parameters by the energy fractions
$x_i$, $x_3$ and the 2-vectors ${\mathbf k}_i$ and ${\mathbf k}$,
the kinematic invariants appearing in the upper vertex take the
following form:
\begin{eqnarray}
  \label{eq:5}
  a_i&\equiv&-(p_1-k_i)^2+m^2=
     \frac{1}{x_i}(m^2x_i^2+{\mathbf k}_i^2) \,,
\nonumber
\\
  b_i&\equiv&(p_3+k_i)^2-m^2=\frac{1}{x_i x_3}(m^2 x_i^2+
  {\mathbf r}_i^2) \,,
\nonumber
\\
  a&\equiv&-(p_1-k_1-k_2)^2+m^2=a_1+ a_2 -\frac{1}{x_1x_2}
  \left(x_1{\mathbf{k}}_2 -x_2{\mathbf{k}}_1 \right)^2
\nonumber
\\
  &=& \frac{1}{x_1x_2}\biggl[x_1x_2(1-x_3)m^2
  +x_2(1-x_2){\mathbf k}_1^2 +x_1(1-x_1){\mathbf k}_2^2
  +2x_1x_2{\mathbf k}_1 {\mathbf k}_2\biggr] \,,
\nonumber
\\
  b&\equiv&(p_3+k_1+k_2)^2-m^2=b_1+b_2+\frac{1}{x_1x_2}
  \left(x_1{\mathbf{k}}_2 -x_2{\mathbf{k}}_1 \right)^2
\\
  &=&\frac{1}{x_1x_2x_3}\biggl[x_1x_2(1-x_3)m^2+x_2(1-x_2)
  ({\mathbf {k}}_1+x_1{\mathbf {k}})^2
\nonumber
\\
  && +x_1(1-x_1)({\mathbf {k}}_2+x_2 {\mathbf {k}})^2
  +2x_1x_2 ({\mathbf {k}}_1+x_1 {\mathbf {k}})
  ({\mathbf {k}}_2+x_2 {\mathbf {k}}) \biggr] \,,
\nonumber
\end{eqnarray}
where
\begin{eqnarray}
  {\mathbf r}_1=x_1({\mathbf k}_2+{\mathbf k})+(1-x_2){\mathbf k}_1
  \,, \quad
  {\mathbf r}_2=x_2({\mathbf k}_1+{\mathbf k})+(1-x_1){\mathbf k}_2
  \label{r1r2}
\end{eqnarray}
or
\begin{eqnarray} \label{ri}
  {\mathbf r}_i=x_3{\mathbf k}_i -  x_i{\mathbf p}_3\,.
\end{eqnarray}
Note that $b= - {\mathbf{k}}^2 - s \alpha_k$ and $b\to a/x_3$
in the limit ${\mathbf{k}}\to 0$.

Using the gauge invariance of an amplitude containing photons we can
replace the polarization vector $e_i$ of a photon $i$\footnote{As usual,
$e_i$ ($e_i^*$) are used for incoming (outgoing) photons.} with
external momentum $k_i$ by $e_i + \zeta_i k_i$.
The arbitrary parameter $\zeta_i$ is chosen in such a way that the
polarization vector does not contain a Sudakov projection onto the
light--like vector $p$:
\begin{eqnarray}
  e_i\equiv e^{(\lambda_i)}(k_i)= \alpha_{e_i}
  p' + e_\perp^{(\lambda_i)} \,.
  \label{polvector}
\end{eqnarray}
The transverse component $e_\perp$ which does not depend on the
4--momentum of the photon $k_i$ is chosen as
\begin{eqnarray}
  \label{eq:9}
  e^{(\lambda_i)}_{\perp}=- \frac{\lambda_i}{\sqrt{2}}
  (0, 1, {\mathrm{i}}\, \lambda_i, 0) \,, \quad
  e^{(\lambda_i)*}_{\perp}=- e^{(-\lambda_i)}_{\perp} \,.
\end{eqnarray}
The Sudakov parameters $\alpha_{e_i}$ are found from the
conditions $e_ik_i=0$:
\begin{eqnarray}
  \label{eq:10}
  s\alpha_{e_i}=\frac{2 {\mathbf {e}}^{(\lambda_i)}
  {\mathbf {k}}_i}{x_i}=
  \frac{\sqrt{2}}{x_i}(-\delta_{\lambda_i,1}\kappa_i+
  \delta_{\lambda_i,-1}\kappa_i^*)\,.
\end{eqnarray}
Here $\kappa_i$ and $\kappa_i^*$ are the circular components of the
vector ${\mathbf{k}}_i$ [compare Eq. (\ref{circular})].

%%%%%%%%%%%%%%%%%%%%%%%%%%%%%%%%%%%%%%%%%%%%%%%%%%%%%%%%%%%%%%%%%%%%%%%%%

\subsection{Helicity amplitudes for the upper vertex factor
$J_{\mathrm{ up}}(e^- \gamma^* \to e^- \gamma \gamma)$}

Now all ingredients are prepared to calculate the helicity
amplitudes for the upper vertex factor. We will use the following
notation indicating the helicity states of the electron
$\lambda_{ei}$ and of the photons $\lambda_i$ explicitly:
\begin{eqnarray}
  J_{\mathrm{up}}\equiv \frac{\sqrt{2} (4 \pi \alpha)^{3/2}}{s}
  \, {\cal{M}}_{\lambda_{e1}\lambda_{e3}}^{\lambda_1\,\lambda_2} \,.
  \label{matelemup}
\end{eqnarray}
The matrix element ${\cal{M}}$ can be represented by the
Feynman diagrams
\begin{figure}[!htb]
  \begin{minipage}{4.5cm}
    \begin{center}
       \epsfig{file=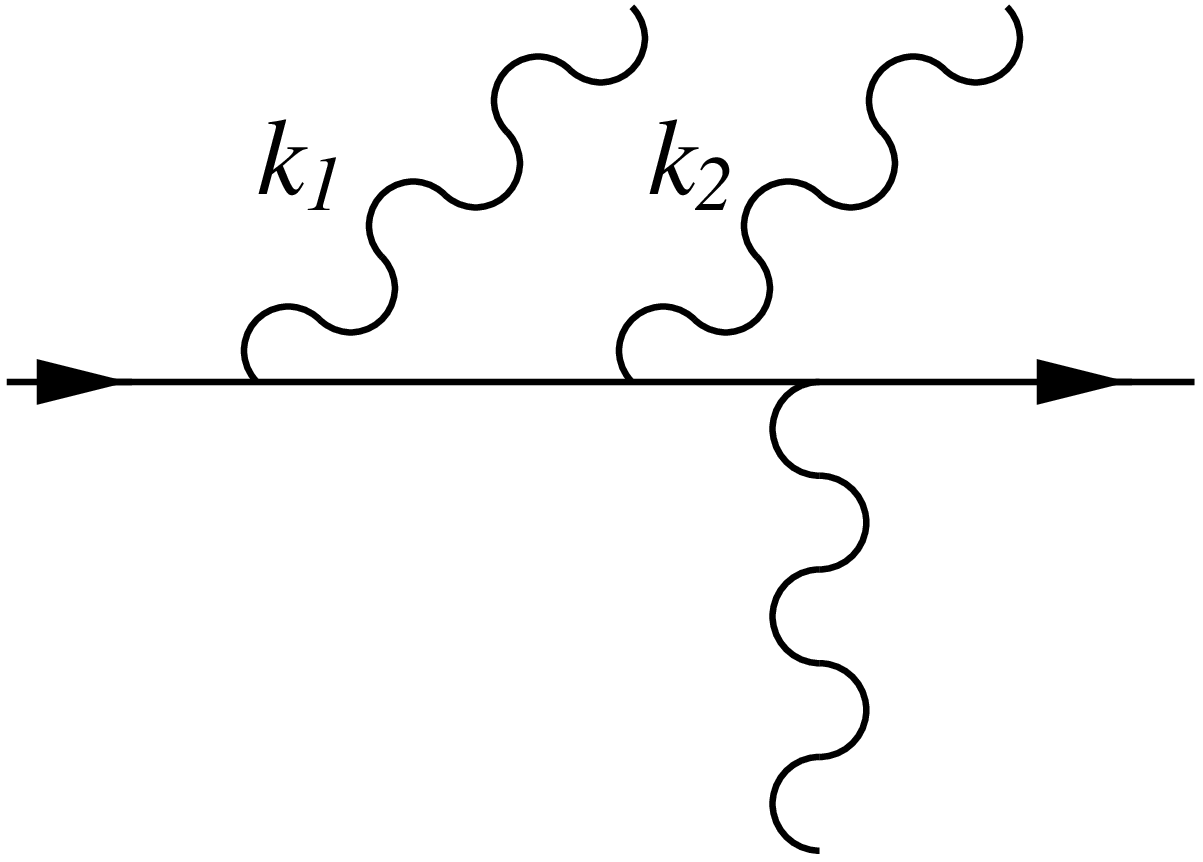,height=40mm,angle=270}
    \end{center}
  \end{minipage}
  \begin{minipage}{4.5cm}
    \begin{center}
       \epsfig{file=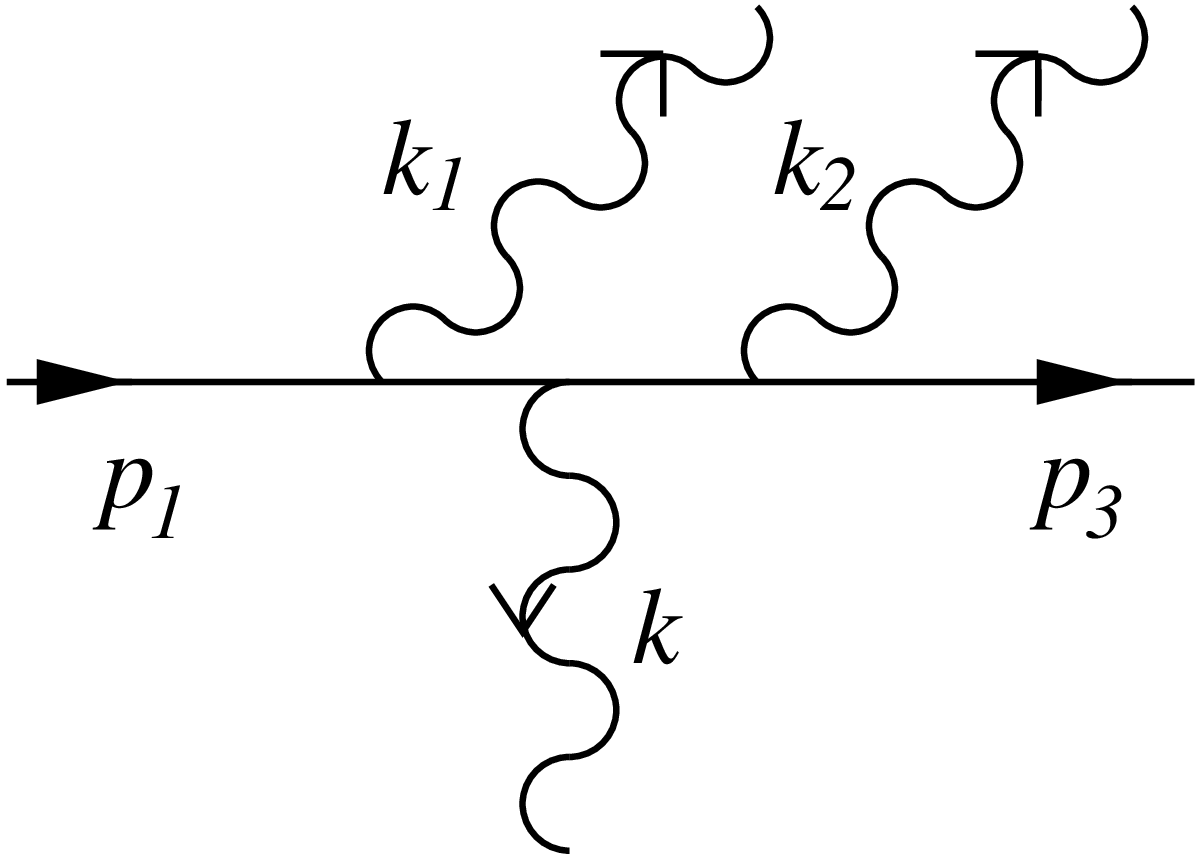,height=40mm,angle=270}
    \end{center}
  \end{minipage}
  \begin{minipage}{4.5cm}
    \begin{center}
       \epsfig{file=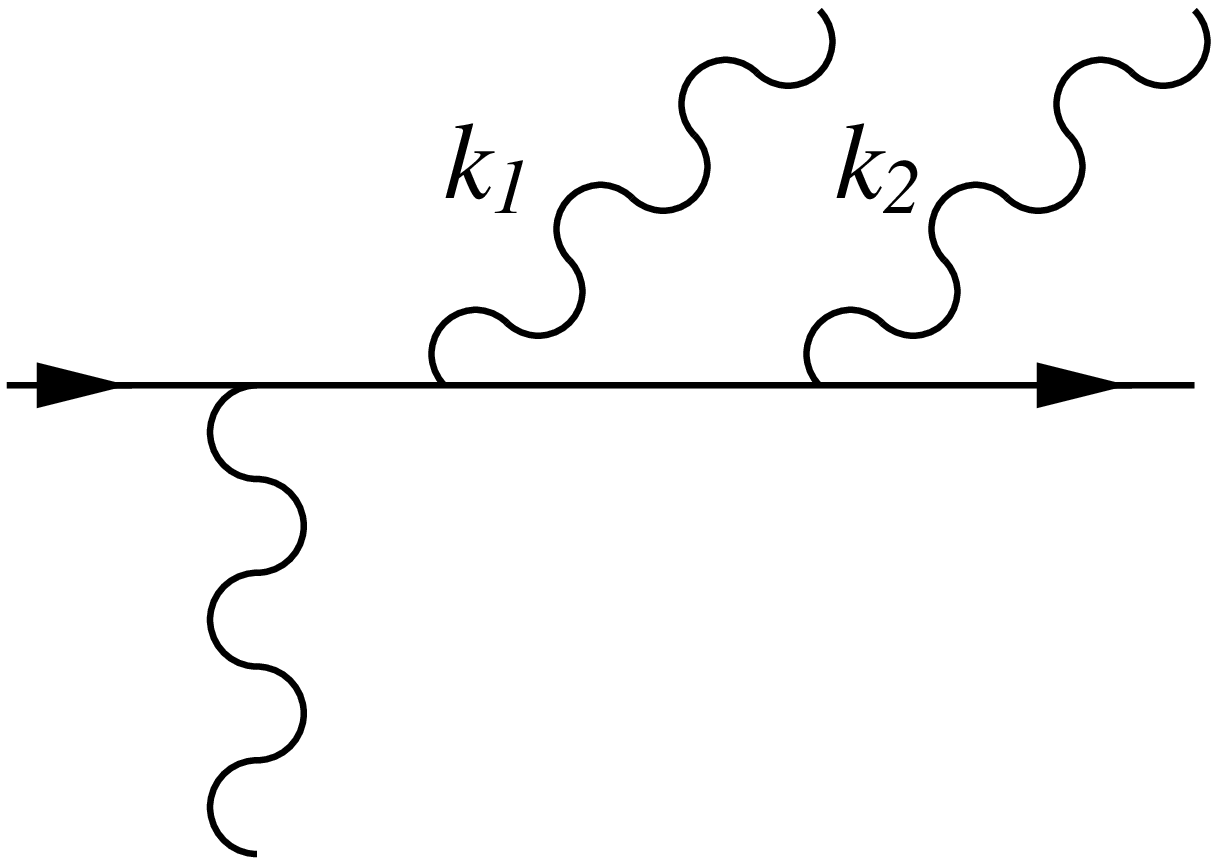,width=40mm,angle=0}
    \end{center}
  \end{minipage}
  \caption{Feynman diagrams for $ e^- \gamma^* \to e^- \gamma\gamma$,
           diagrams with $k_1\leftrightarrow k_2$ have to be added.}
  \label{fig:3}
\end{figure}
shown in Fig.~\ref{fig:3}. The amplitudes corresponding to that
figure are
\begin{eqnarray}
  \label{eq:12}
  {\cal{M}}=(1+{\cal{P}}_{12})Q \,, \qquad
  Q={\cal{M}}_1+{\cal{M}}_2+{\cal{M}}_3\,,
\end{eqnarray}
where
\begin{eqnarray}
  \label{amplitudes}
  {\cal{M}}_1&=&\frac{1}{a_1 a}\bar u_3\hat p'(\hat p_1-\hat k_1
  -\hat k_2+m)\hat e_2^*(\hat p_1-\hat k_1+m)
  \hat e_1^*u_1\,,
\nonumber
\\
  {\cal{M}}_2&=&-\frac{1}{a_1 b_2}\bar u_3
  \hat e_2^*(\hat p_1-\hat k_1-\hat k+m)\hat p'(\hat p_1-\hat k_1+m)
  \hat e_1^* u_1\,,
\\
  {\cal{M}}_3&=&\frac{1}{b_2 b}\bar u_3
  \hat e_2^*(\hat p_1-\hat k_1-\hat k+m)
  \hat e_1^*(\hat p_1-\hat k+m) \hat p 'u_1 \,.
\nonumber
\end{eqnarray}
The permutation operator ${\cal{P}}_{12}$ for the photons is defined as
\begin{eqnarray*}
{\cal{P}}_{12}f(k_1,e_1;k_2,e_2)=
f(k_2,e_2;k_1,e_1)\,, \qquad {\cal{P}}_{12}^2 =1. \nonumber
\end{eqnarray*}

The quantity $Q$ is gauge invariant with respect to the virtual photon $k$
since all permutations of this photon are taken into account.
Therefore $Q$ is proportional to $k_\perp$ in the limit
$k_\perp \to 0$. Indeed, using the relations
\begin{eqnarray}
  Q= p'_\mu Q^\mu \,, \ \ \ k_\mu Q^\mu=
  (\alpha_k p' +\beta_k p +k_\perp)_\mu Q^\mu =0\,,
\end{eqnarray}
we immediately obtain (neglecting the small contribution
$\beta_k p_\mu Q^\mu\sim 1/s$)
\begin{eqnarray} \label{eq:13}
  Q= -\frac{k_{\perp\mu}}{\alpha_k} Q^\mu \,.
\end{eqnarray}

Now we transform the quantities ${\cal M}_j$ to such a form that in
their sum $Q$ this noticed low $k_\perp$ behaviour is present
explicitly. The reason is that in that case all individual large
(compared to $k_\perp$) contributions are cancelled. The first step
is to use the Dirac equations $\hat p_1 u_1 =m u_1$,
$\bar u_3 \hat p_3 = m \bar u_3$ and to rearrange
the amplitudes ${\cal{M}}_j$ of Eq.~(\ref{amplitudes}) as follows:
\begin{eqnarray}
  \label{eq:14}
  {\cal{M}}_1&=& \bar u_3 \Biggl\{
  \frac{sx_3}{a_1 a}  \hat e_2^*(\hat p_1-\hat k_1+m) \hat e_1^*  +
  \frac{1}{a_1 a}  \hat p'\hat k \hat e_2^*
  (\hat p_1-\hat k_1+m) \hat e_1^* \Biggr\} u_1\,,
\nonumber
\\
  {\cal{M}}_2&=& \bar u_3 \Biggl\{
  -\frac{s(1-x_1)}{a_1 b_2} \hat e_2^*
  (\hat p_1-\hat k_1+m) \hat e_1^*
  - \frac{1}{b_2} \hat e_2^* \hat p' \hat e_1^*
\nonumber
\\
  && +\frac{1}{a_1 b_2} \hat e_2^* \hat k \hat p'
  (\hat p_1- \hat k_1+m) \hat e_1^* \Biggr\} u_1 \,,
\\
  {\cal{M}}_3&=& \bar u_3 \Biggl\{
  \frac{s}{b_2 b}  \hat e_2^* (\hat p_1-\hat k_1+m) \hat e_1^*
  -\frac{s}{b_2 b}  \hat e_2^* \hat k \hat e_1^*
\nonumber
\\
  && -\frac{1}{b_2 b}  \hat e_2^*(\hat p_3+\hat k_2+m)
  \hat e_1^* \hat k \hat p' \Biggr\} u_1 \,.
\nonumber
\end{eqnarray}
{}From these formulae we observe that the last terms in
${\cal{M}}_1, {\cal{M}}_2, {\cal{M}}_3$ are proportional to
$k_{\perp}$ up to terms of the order of (\ref{accuracy}):
\begin{eqnarray}
  \label{eq:15}
  \hat p'\hat k=\hat p' (\alpha_k \hat p' +
  \beta_k \hat p + \hat k_\perp) =
  \hat p'\hat k_\perp= - \hat k \hat p' \,.
\end{eqnarray}
As a next step we note that the sum of the first three terms
in Eqs.~(\ref{eq:14}) is also proportional to $k_\perp$ since
[see Eqs.~(\ref{eq:5})]
\begin{eqnarray}
  \label{eq:16}
  A \equiv \frac{x_3}{a_1 a}-\frac{1-x_1}{a_1 b_2}+\frac{1}{b_2 b} \,,
  \qquad A |_{k_{\perp}\to 0}=0\,.
\end{eqnarray}
Finally we consider the sum of the second terms of the quantities
${\cal{M}}_2, {\cal{M}}_3$ given in Eqs.~(\ref{eq:14}). Using the
relations~(\ref{eq:5}) and (\ref{invmass}) one gets
\begin{eqnarray}
  \label{eq:17}
  -\frac{\hat p'}{b_2}-\frac{s(\alpha_k \hat p'+
  \hat k_{\perp})}{b_2 b}=
  -\frac{s\hat k_{\perp}}{b_2 b}+
  \frac{\hat p' {\mathbf {k}}^2}{b_2 b} \,.
\end{eqnarray}
Therefore, from Eqs.~(\ref{eq:15}),~(\ref{eq:16}),~(\ref{eq:17})
it is clearly seen that the discussed property (\ref{eq:13})
\begin{eqnarray*}
  \left({\cal{M}}_1+{\cal{M}}_2+{\cal{M}}_3 \right)
  \vert_{k_{\perp}\to 0}=0
\end{eqnarray*}
is obviously satisfied and
the quantity
$Q=\sum_{j=1}^3 {\cal{M}}_j$ is a sum of terms explicitly
proportional to $k_{\perp}$:
\begin{eqnarray}
  \label{eq:18}
  Q&=&\bar u_3\biggl\{A s \,\hat e_2^*
  (\hat p_1-\hat k_1+m) \hat e_1^*
  +\frac{1}{a_1 a} \hat p' \hat k_{\perp} \hat e_2^*
  (\hat p_1-\hat k_1+m) \hat e_1^*
\nonumber
\\
  && -\frac{{\mathbf {k}}^2} {b_2 b} \hat e_2^* \hat e_1^* \hat p'
  - \frac{s}{b_2 b} \hat e_2^* \hat k_{\perp} \hat e_1^*
\\
  && +\frac{1}{a_1 b_2} \hat e_2^* \hat k_{\perp} \hat p'
  (\hat p_1-\hat k_1+m) \hat e_1^* -\frac{1}{b_2 b}
  \hat e_2^* ( \hat p_3+\hat k_2+m) \hat e_1^* \hat k_{\perp}
  \hat p'\biggr\}u_1\,.
\nonumber
\end{eqnarray}
The subsequent calculations require to determine several bilinear
spinor combinations which are listed in Appendix \ref{appa}.

As a result, we obtain the complete expression for the amplitude of
double forward bremsstrahlung in a jet kinematics
\begin{eqnarray} \label{fullamp}
  M(e^- e^+ \to e^- \gamma \gamma + e^+) =
  \frac {32 \pi^2 \alpha^2 }{k^2} \,
  {\cal{M}}_{\lambda_{e1}\lambda_{e3}}^{\lambda_1\;\lambda_2}\,
  {\mathrm{e}}^{-{\mathrm{i}} \, \lambda_{\bar{e}4} \, \varphi_4}
  \,\delta_{\lambda_{\bar{e}2},\lambda_{\bar{e}4}} \,.
\end{eqnarray}
All helicity amplitudes
${\cal{M}}_{\lambda_{e1}\lambda_{e3}}^{\lambda_1\;\lambda_2}$
are expressed via the complex circular components
$\kappa_i$, $\kappa$ and $\rho_i$ of the vectors
${\mathbf{k}}_i$, $\mathbf{k}$ and ${\mathbf{r}}_i$
[see Eqs.~(\ref{circular}) and (\ref{r1r2})]
\begin{eqnarray}
  \label{eq:20}
  \kappa_i=k_{ix}+ {\mathrm{i}} \, k_{iy} \,, \quad
  \kappa=k_x+ {\mathrm{i}} \, k_y \,, \quad
  \rho_i=r_{ix}+{\mathrm{i}}\, r_{iy}\,, \quad i=1,2\,.
\end{eqnarray}
The spin non--flip amplitudes (with
$\lambda_{e1}=\lambda_{e3}=+1/2 \equiv +$) being proportional to
the factor
\begin{eqnarray} \label{S}
  S_{\lambda_{e3}}= s \sqrt{x_3} {\mathrm{e}}^{{\mathrm{i}} \,
  \varphi_3 \lambda_{e3}}\,,
\end{eqnarray}
are equal to\footnote{To clarify the notation we stress that in
Eqs.~(\ref{nonflip}) and (\ref{flip}) the operator ${\cal{P}}_{12}$
simply changes the indices $1\leftrightarrow 2$, for instance
\begin{eqnarray*}
{\cal{P}}_{12}{\cal{M}}^{+-}_{+-}=2mS_-({\cal{P}}_{12}F + G),\quad
{\cal{P}}_{12}F = x_2\left[({\cal{P}}_{12}A)\frac{\rho_1}{x_1x_3}
- \frac{\kappa}{a_2a}\right],
\end{eqnarray*}
with
\begin{eqnarray*}
{\cal{P}}_{12}A = \frac{x_3}{a_2a} - \frac{1-x_2}{a_2b_1} + \frac{1}{b_1b}.
\end{eqnarray*}
}
\begin{eqnarray}
  \label{nonflip}
  {\cal{M}}^{++}_{++}&=&
  2 S_+ (1+{\cal{P}}_{12})\Biggl\{
  A \frac{\kappa_1^* \rho_2^*}{x_1 x_2 x_3}
  -\frac{\kappa_1^*\kappa^*}{x_1 a_1 a }
  + \frac{\kappa^* \rho_2^*}{x_2 x_3  b_2 b} \Biggr\} \,,
\nonumber
\\
  {\cal{M}}^{--}_{++}&=&
  x_3 S_+ \left({\cal{M}}^{++}_{++}\over S_+\right)^* \,,
\nonumber
\\
  {\cal{M}}^{+-}_{++}&=&-2 S_+ \left( C + {\cal{P}}_{12} D\right)\,,
\nonumber
\\
 {\cal{M}}^{-+}_{++}&=&{\cal{P}}_{12} {\cal{M}}^{+-}_{++} \,,
\\
  C&=& A \left(
  \frac{\kappa_1^* \kappa_2}{x_1x_2}-\frac{x_1}{x_3} m^2 \right)
  -\frac{\kappa_1^* \kappa}{x_1 a_1 b_2 }+
  \frac{x_1 x_2 \kappa^* \kappa + x_1 \kappa^* \kappa_2
  +x_2 \kappa_1^* \kappa}{x_1 x_2  b_2 b } \,,
\nonumber
\\
  D&=& (1-x_1) \Biggl(
  A \frac{\kappa_1 \rho_2^*}{x_1x_2x_3}
  -\frac{\kappa_1\kappa^*}{x_1 a_1 a }
  +\frac{\kappa\rho_2^*}{x_2 x_3  b_2 b }\Biggr) \,.
\nonumber
\end{eqnarray}
Another convenient form of $C$ useful for the soft photon limit
(see Section~\ref{softphotonlimit})
and for the later crossing (Section~\ref{crossingsection}) is
\begin{eqnarray}
  \label{alternC}
  C&=& \frac{1-x_2}{1-x_1} D^* + \Delta,
\\
  \Delta&=& -\frac{1}{a}+\frac{1}{b}
  +\frac{x_1x_2 m^2 +\kappa_1^* \rho_2}{x_3 a_1 b_2}
  +\frac{\kappa_1^* (x_2 \kappa_1 - x_1 \kappa_2)}{x_1 a_1 a}
  -\frac{(x_2 \kappa_1^*  -x_1 \kappa_2^*) \rho_2}{x_2 x_3 b_2 b} \,.
\nonumber
\end{eqnarray}
Note that in the limit of a soft (second) photon the function
$D\propto 1/x_2$ whereas $\Delta$ remains finite.

For the spin flip amplitudes (with $\lambda_{e1}=-\lambda_{e3}=+1/2$)
proportional to $ m S_-$ we obtain
\begin{eqnarray} \label{flip}
  {\cal{M}}^{++}_{+-}&=&2 m S_- (1+{\cal{P}}_{12})
  \Biggl\{
  -\frac{1-x_3}{x_2} G^* + x_1 \left[ A \left( \frac{\kappa_1^*}{x_1}
  -\frac{\kappa_2^*}{x_2} \right) + \frac{\kappa^*}{a_1 b_2} \right]
  \Biggr\}\,,
\nonumber
\\
  {\cal{M}}^{--}_{+-}&=&0 \,,
\nonumber
\\
  {\cal{M}}^{+-}_{+-}&=& 2 m S_- \left( F + {\cal{P}}_{12} G\right) \,,
\\
  {\cal{M}}^{- +}_{+-}&=& {\cal{P}}_{12} {\cal{M}}^{+-}_{+-} \,,
\nonumber
\\
  F& =&x_1
  \Biggl( A \frac{\rho_2}{x_2 x_3} -\frac{\kappa}{a_1 a}\Biggr)\,,
  \quad
  G = \frac{x_2}{x_3}  \Biggl(
  A \frac{\kappa_1}{x_1} + \frac{\kappa}{b_2 b}\Biggr) \,.
\nonumber
\end{eqnarray}

The remaining amplitudes  with  $\lambda_{e1}=-1/2$ can be obtained
by applying the parity conservation relation
\begin{eqnarray}
  {\cal{M}}_{-\lambda_{e1},-\lambda_{e3}}^{-\lambda_1,\;\,-\lambda_2}
  = - (-1)^{\lambda_{e1}+ \lambda_{e3}} \left(
  {\cal{M}}_{\lambda_{e1},\lambda_{e3}}^{\lambda_1, \;\, \lambda_2}
  \right)^* \,.
  \label{parity}
\end{eqnarray}
This rule follows from the transformation properties of the electron
and photon wave functions under space inversion [see \S 16 in
Ref.~\cite{BLP} and Eq.~(\ref{eq:9})]:
\begin{eqnarray}
  u_i\equiv u_{{\mathbf{p}}_i}^{(\lambda_{ei})} \to
  {\mathrm{i}} \, (-1)^{s- \lambda_{ei}}
  u_{{\mathbf{p}}_i}^{(-\lambda_{ei})} \,, \quad
  {\mathbf{pe}}^{(\lambda_i)} \to
  {\mathbf{pe}}^{(-\lambda_i)}= -
  {\mathbf{pe}}^{(\lambda_i)*}\,,
\end{eqnarray}
where $s=1/2$ and ${\mathbf{p}}$ is an arbitrary 2--vector.
These properties give an additional factor
$(-1)^{n_\gamma + 2 s - \lambda_{e1} -\lambda_{e3}}$ where
$n_\gamma=2$ is the number of real photons. Therefore, under
replacement (\ref{parity}) we have to take the complex conjugates of
the original helicity amplitudes with sign plus for the spin non-flip
amplitudes (\ref{nonflip}) and sign minus for the spin flip ones
(\ref{flip}).

All helicity amplitudes in Eqs.~(\ref{nonflip}) and (\ref{flip})
are explicitly proportional to $\kappa$ or to the function $A$.
Therefore, they vanish $\propto |{\mathbf{k}}|$ in the limit
$|{\mathbf{k}}| \to 0$. It is clear (although not obvious) that also
$\Delta$ in Eq.~(\ref{alternC}) vanishes in that limit
since both $C$ and $D$ in (\ref{nonflip}) tend to zero.

We note the explicit Bose symmetry  between the two photons
in the amplitudes~(\ref{nonflip}) and (\ref{flip}):
\begin{eqnarray}
  {\cal{M}}^{\lambda_1 \;\, \lambda_2}_{\lambda_{e1} \lambda_{e3}}
  (\kappa_1,x_1;\kappa_2,x_2)=
  {\cal{M}}^{\lambda_2 \;\, \lambda_1}_{\lambda_{e1} \lambda_{e3}}
  (\kappa_2,x_2;\kappa_1,x_1) \,.
\end{eqnarray}
Since the spin--flip amplitudes are proportional to the lepton mass
they are negligible compared to the spin non--flip ones for not
too small scattering angles
\begin{eqnarray}
  \frac{m}{x_{1,2,3}E_1} \ll \theta_{1,2,3} \ll 1 \,.
  \label{langle}
\end{eqnarray}
We observe that in our kinematics the transitions with the largest
change of helicities between initial and final particles
$ \vert \lambda_{e1}- \lambda_{e3}-\lambda_1-\lambda_2 \vert=3$
are forbidden, in our case the amplitude ${\cal{M}}_{+-}^{--}=0$.
This is in agreement with the vanishing amplitude for single photon
bremsstrahlung with the largest change of helicities
$ \vert \lambda_{e1} - \lambda_{e3} - \lambda_1 \vert=2$,
see later Eq.~(\ref{jup}).

%%%%%%%%%%%%%%%%%%%%%%%%%%%%%%%%%%%%%%%%%%%%%%%%%%%%%%%%%%%%%%%%%%%%%%%

\section{Limiting cases  of soft and hard final particles}

\subsection{The soft photon limit}
\label{softphotonlimit}

A useful check of the results obtained is the limiting case in which
one of the real photons (belonging to the upper block
$J_{\mathrm{up}}$) becomes soft. For definiteness let us suppose
that this is the second photon which means that $ x_2 \ll1$,
$k_2 \to 0$, but ${\mathbf{k}}_2/x_2$ remains finite.

It is known that in this limit the upper block factorizes and can be
presented in the form
\begin{eqnarray}
  J_{\mathrm{up}}= \sqrt{4 \pi \alpha} \,
  J_{\mathrm{up}}(e^-_{\lambda_{e1}} \gamma^*
  \to e^-_{\lambda_{e3}} \gamma_{\lambda_1}) \, A_2 \, ,
\end{eqnarray}
where
\begin{eqnarray}
  A_2= \left(\frac{p_1}{p_1 k_2}-\frac{p_3}{p_3 k_2} \right) e_2^{*}
\end{eqnarray}
is the factor of the accompanying classical emission of the second
photon. This factor is easily transformed to
\begin{eqnarray}
  A_2=  \frac{2}{x_2}
  \left( \frac{{\mathbf {k}}_2}{a_2}-\frac{{\mathbf {r}}_2}{b_2}
  \right) {\mathbf {e}}^{(\lambda_2)*} \,.
\end{eqnarray}
The vertex factor $J_{\mathrm{up}}(e^-_{\lambda_{e1}} \gamma^*
\to e^-_{\lambda_{e3}} \gamma_{\lambda_1})$ can be found
in Ref.~\cite{KSSS}:
\begin{eqnarray}
  \label{jup}
  &&J_{\mathrm{up}}(e^-_{\lambda_{e1}} \gamma^*
  \to e^-_{\lambda_{e3}} \gamma_{\lambda_1})=
  8 \pi \alpha \,\sqrt{1-x_1} \,{\mathrm e}^{ {\mathrm i} \,
  \varphi_3 \lambda_{e3}}
\\
  &&\times
  \left\{
   \sqrt{2} \, {\mathbf{Q}}_1 {\mathbf{e}}^{(\lambda_1)*} \,
   \frac{ 1- x_1 \delta_{ \lambda_1, -2 \lambda_{e1}} } {x_1}
   \delta_{\lambda_{e1},\lambda_{e3}}
   + m R_1 \delta_{\lambda_1, 2 \lambda_{e1}}
   \delta_{\lambda_{e1},-\lambda_{e3}}
  \right\} \,.
\nonumber
\end{eqnarray}
Here we use the notations (${\mathbf{q}}_1= {\mathbf{k}}_1/x_1$)
\begin{eqnarray} \label{QR}
  {\mathbf {Q}}_1&=&\frac{{\mathbf {k}}_1}{a_1}-
  \frac{{\mathbf {r}}_1}{x_3 b_1}
  = \frac{{\mathbf{q}}_1}{m^2+{\mathbf{q}}_1^2}
  -\frac{{\mathbf{k}}+{\mathbf{q}}_1}{m^2+({\mathbf{k}}
  +{\mathbf{q}}_1)^2} \,,
\\
  R_1&=&\frac{x_1}{a_1}- \frac{x_1}{x_3 b_1}
  = \frac{1}{m^2+{\mathbf{q}}_1^2}
  - \frac{1}{m^2+({\mathbf{k}}+{\mathbf{q}}_1)^2} \,.
\nonumber
\end{eqnarray}

Therefore, in the soft photon limit the spin non--flip and flip
amplitudes are
\begin{eqnarray} \label{softphoton}
{\cal{M}}_{\lambda_{e1} \lambda_{e1}}^{\lambda_1 \;\, \lambda_2}&=&
  2 S_{\lambda_{e1}} {\mathbf{Q}}_1 {\mathbf{e}}^{(\lambda_1)*}
  \frac{1-x_1 \delta_{\lambda_1,- 2 \lambda_{e1}}}{x_1} A_2 \,, \\
{\cal{M}}_{\lambda_{e1},-\lambda_{e1}}
  ^{\lambda_1 \ \ \;\lambda_2}&=&
  \sqrt{2} m \, S_{-\lambda_{e1}} \, R_1 \, \delta_{\lambda_1,2
  \lambda_{e1}}A_2 \,.
\nonumber
\end{eqnarray}
These expressions coincide with those from the amplitudes
(\ref{nonflip}), (\ref{flip}) in the limit of a soft (second) photon
when
\begin{eqnarray}
  &&a \to a_1 \,, \quad
  b \to b_1 \,, \quad
  \rho_1 \to \kappa_1 + x_1 \kappa \,,
\\
  && A \to -\frac{x_3}{a_1 b_2} + \frac{1}{b_1 b_2} \,,  \quad
  {\cal P}_{12} A \to \frac{x_3}{a_1 a_2}- \frac{1}{b_1 a_2} \,.
\nonumber
\end{eqnarray}
This can be easily shown for most of the helicity amplitudes given
in Eqs. (\ref{nonflip}), (\ref{flip}), for the amplitude
${\cal{M}}_{++}^{+-}$ we use the form (\ref{alternC}) for $C$ and
neglect $\Delta$ in this limit.

%%%%%%%%%%%%%%%%%%%%%%%%%%%%%%%%%%%%%%%%%%%%%%%%%%%%%%%%%%%%%%%%%%%

\subsection{The soft electron limit $m\ll E_3 \ll E_1$}

For this limit we assume that the final electron has an energy $E_3$
much smaller than that of the initial lepton $E_1$, but the lepton
remains ultrarelativistic. This leads to the condition $x_3 \ll 1$
whereas the value of the 2--vector ${\mathbf{p}}_3$ is comparable to
those of the other 2--vectors. In that limit the helicity amplitudes
vanish with $\sqrt{x_3}$ due to the factor $S_{\lambda_{e3}}$
[see Eq.~(\ref{S})]  and the cross section tends to zero with $x_3$.
Besides of this obvious factor the remaining dominant part of the
amplitude (staying finite in that limit) can be found.

The kinematic invariants $a$, $b$  and $b_i$  are simplified to
\begin{eqnarray}
  &&   a  = m^2 +({\mathbf{k}}_1+{\mathbf{k}}_2)^2  \,, \quad
  x_3 b   = m^2 + ({\mathbf{k}}+{\mathbf{k}}_1+ {\mathbf{k}}_2)^2
          = m^2 + {\mathbf{p}}_3^2 \,,
\nonumber
\\
  &&b_i   = x_i b \,.
\end{eqnarray}
Additionally we have
\begin{eqnarray}
  A \to \frac{x_3}{a_1} \left( \frac{1}{a}-\frac{1}{x_3 b}\right)
  \,, \quad
  {\cal P}_{12}A \to \frac{x_3}{a_2}
  \left( \frac{1}{a}-\frac{1}{x_3 b}\right) \,, \quad
  \rho_i \to -x_i \pi_3 \,.
\end{eqnarray}
{}From Eqs.~(\ref{nonflip}) and (\ref{flip}) we derive
\begin{eqnarray}
  &&{\cal{M}}^{++}_{++}= 2 S_+
  \left( \frac{\kappa_1^*}{x_1 a_1}+\frac{\kappa_2^*}{x_2 a_2} \right)
  \left( \frac{\kappa_1^*+\kappa_2^*}{a}+\frac{\pi_3^*}{x_3 b} \right)   \,,
\nonumber
\\
  &&{\cal{M}}^{--}_{++}=0 \,,
\nonumber
\\
  &&{\cal{M}}^{+-}_{++}= -2 x_1 S_+
  \left\{ \frac{\kappa_2}{x_2 a_2}
  \left( \frac{\kappa_1^*+\kappa_2^*}{a}+ \frac{\pi_3^*}{x_3 b}\right)
  - m^2 \frac{1}{a_1} \left(\frac{1}{a}-\frac{1}{x_3 b}\right)
  \right\} \,,
\nonumber
\\
  &&{\cal{M}}^{-+}_{++}={\cal{P}}_{12} {\cal{M}}^{+-}_{++} \,,
\\
  &&{\cal{M}}^{++}_{+-}=
   - 2 m S_- \left(\frac{1}{a}-\frac{1}{x_3 b} \right)
  \left(\frac{\kappa_1^*}{x_1 a_1}+\frac{\kappa_2^*}{x_2 a_2}
  \right)\,,
\nonumber
\\
  &&{\cal{M}}^{+-}_{+-}= 2 m x_1 S_-
  \left\{ \frac{1}{a_1}
   \left(  \frac{\kappa_1+\kappa_2}{a}+\frac{\pi_3}{x_3 b} \right)
   + \left(\frac{1}{a}-\frac{1}{x_3b} \right)
    \frac{\kappa_2}{x_2 a_2}
   \right\} \,,
\nonumber
\\
  &&{\cal{M}}^{-+}_{+-}={\cal{P}}_{12} {\cal{M}}^{+-}_{+-} \,,
\nonumber
\\
  &&{\cal{M}}^{--}_{+-}=0\,.
\nonumber
\end{eqnarray}

Introducing the notations
\begin{eqnarray}
  {\mathbf{Q}}_e =
  \frac{{\mathbf{k}}_1+{\mathbf{k}}_2}{a}+
  \frac{{\mathbf{p}}_3}{x_3 b}
  \,, \quad
  R_e =\frac{1}{a}-\frac{1}{x_3 b}\,,
\end{eqnarray}
the matrix elements in the soft electron limit can be written as
\begin{eqnarray}
  \label{softelectron}
  &&{\cal{M}}_{\lambda_{e1}\lambda_{e3}}^{\lambda_1\;\, \lambda_2}=
  \sqrt{x_3} s \,{\mathrm{e}}^{{\mathrm{i}} \,
  \varphi_3 \lambda_{e3}  }
  \left( 1 + {\cal{P}}_{12} \right)
  \Biggl(
       2 {\mathbf{Q}}_e{\mathbf{e}}^{(\lambda_2)*} \,
         \delta_{\lambda_2,2\lambda_{e3}}
       + \sqrt{2} m R_e \delta_{\lambda_2,-2\lambda_{e3}}
  \Biggr)
\nonumber
\\
  &&\times \frac{1}{a_1}
  \Biggl(
       \frac{1-x_1\delta_{\lambda_1,-\lambda_2} }{x_1}
      2 {\mathbf{k}}_1{\mathbf{e}}^{(\lambda_1)*} \,
        \delta_{\lambda_2,2\lambda_{e1}}
      + \sqrt{2}m x_1 \delta_{\lambda_1,-\lambda_2}
      \delta_{\lambda_2,-2\lambda_{e1}}
  \Biggr) \,.
\end{eqnarray}

Considering additionally a soft (second) photon we arrive at the
case (\ref{hardphoton}) discussed in Section~\ref{hard}.

%%%%%%%%%%%%%%%%%%%%%%%%%%%%%%%%%%%%%%%%%%%%%%%%%%%%%

\subsection{The limits of a hard electron or a hard photon}
\label{hard}

In the limit of a hard electron $x_3\to 1$ both photons become
soft. Using Eqs.~(\ref{softphoton}) for that case,  only the
non--flip amplitudes remain:
\begin{eqnarray}
  {\cal{M}}_{\lambda_{e1}\lambda_{e3}}^{\lambda_1 \; \lambda_2}
  &=& s \, A_1 A_2 \,
  {\mathrm{e}}^{{\mathrm{i}} \, \varphi_3 \lambda_{e3}}
  \delta_{\lambda_{e1},\lambda_{e3}} \,,
\\
  A_i = \frac{2}{x_i}
  \left( \frac{{\mathbf {k}}_i}{a_i}-\frac{{\mathbf {r}}_i}{b_i}
  \right)
  {\mathbf {e}}^{(\lambda_i)*}
  &=&
  2 \left(\frac{{\mathbf{k}}_i {\mathbf {e}}^{(\lambda_i)*}}
             {x_i^2 m^2+{\mathbf{k}}_i^2}
  -\frac{ ({\mathbf{k}}_i +x_i
    {\mathbf{k}}){\mathbf {e}}^{(\lambda_i)*}}
    {x_i^2m^2+({\mathbf{k}}_i + x_i {\mathbf{k}})^2}\right)  \,.
\nonumber
\end{eqnarray}

Now we consider the case when the first photon is hard and the soft
electron remains ultrarelativistic: $x_1\to 1$ leading to
$x_{2,3}\ll 1$, $|{\mathbf{r}}_2| /b_2 \ll |{\mathbf{k}}_2|/a_2$
and $ A_2 \to 2{\mathbf{k}}_{2}{\mathbf{e}}^{(\lambda_2)*}/(x_2 a_2)$.
The amplitudes can be derived from Eqs.~(\ref{softphoton}) where
${\mathbf{Q}}_1$ and $R_1$ from Eqs.~(\ref{QR}) have to be taken at
$x_1\to 1$:
\begin{eqnarray}
  \label{hardphoton}
  {\cal{M}}_{\lambda_{e1} \lambda_{e3}}^{\lambda_1 \; \lambda_2}&=&
  \sqrt{x_3} s \, {\mathrm{e}}^{ {\mathrm{i}} \, \varphi_3
  \lambda_{e3}}
  \left( 2  {\mathbf{Q}}_1
  {\mathbf{e}}^{(\lambda_1)*}
  \delta_{\lambda_{e3},\lambda_{e1}}
  +\sqrt{2} m \, R_1 \delta_{-\lambda_{e3},\lambda_{e1}}\right)
\\
  &&\times
  \delta_{\lambda_1,2\lambda_{e1}}\,
  \frac{2}{x_2 a_2} {\mathbf{k}}_2 {\mathbf{e}}^{(\lambda_2)*}
  \,.
\nonumber
\end{eqnarray}
{}From that expression one can see that only those amplitudes remain
in which the initial electron transfers its helicity to the hard photon:
$\lambda_1= 2 \lambda_{e1}$, and they vanish as $\sqrt{x_3}$.

%%%%%%%%%%%%%%%%%%%%%%%%%%%%%%%%%%%%%%%%%%%%%%%%%%%%%%%%%%%%%%%%%%%%%

\section{The cross channel
\boldmath $\gamma e^\pm \to \gamma e^+e^- + e^\pm$ \unboldmath}
\label{crossingsection}

\subsection{Crossing relations and helicity amplitudes}

We consider the reaction
\begin{eqnarray}  \label{crossreaction}
  \gamma(\tilde k_1) + e^\pm(p_2) \to
  \gamma(\tilde k_2) + e^+(p_+) + e^-(p_-) + e^\pm(p_4)\,,
\end{eqnarray}
which is one of the cross channels to the double forward
bremsstrahlung considered before. We denote in the cross channel
the 4--momenta of the initial photon by $\tilde k_1$ with energy
$\tilde \omega_1$ and helicity $\lambda_1$, the 4--momenta
(Euclidean 2--vectors) of the final lepton pair by
$p_\pm$ (${\mathbf{p}}_\pm$) and the final photon by
$\tilde{k}_2$ ($\tilde{\mathbf{k}}_2$)
with helicities $\lambda_\pm$ and $\lambda_2$, respectively.

The amplitude for reaction (\ref{crossreaction}) can be obtained in
the jet kinematics by considering the cross channel of the upper
vertex only:
\begin{eqnarray}
  M(\gamma e^\pm \to \gamma e^+e^- + e^\pm)=
  \frac{\tilde s}{k^2} J_{\mathrm{up}}(\gamma \gamma^* \to \gamma
  e^+e^-) J_{\mathrm{down}}(e^\pm \gamma^* \to e^\pm)\,,
\end{eqnarray}
with
\begin{eqnarray}
  J_{\mathrm{up}}(\gamma \gamma^* \to \gamma e^+e^-)=
  J_{\mathrm{up}}^{\mathrm{cross}}(e^- \gamma^* \to e^-
  \gamma \gamma) \,,\quad \tilde s =2 \tilde k_1 p_2\,.
\end{eqnarray}
The following 4--vector replacements have to be performed:
\begin{eqnarray}
  k_1\to - \tilde k_1 \,, \quad
  k_2\to   \tilde k_2\,, \quad
  p_1\to - p_+ \,, \quad
  p_3\to   p_- \,,
  \label{replace}
\end{eqnarray}
with the remaining 4--vectors $k$, $p_2$ and $p_4$ held intact.

Obviously, the corresponding energy fractions are
\begin{eqnarray}
  y_\pm=\frac{E_\pm}{\tilde \omega_1}\,, \quad
  y_2=\frac{\tilde\omega_2}{\tilde\omega_1} \,, \quad
  y_+ +y_-+y_2=1 \,.
\end{eqnarray}
The Sudakov decomposition is easily performed (using $\tilde k_1$ and $\tilde
p'=p_2 - (m^2/\tilde s)\tilde k_1$ as
light--like vectors instead of $p$ and $p'$, respectively)
\begin{eqnarray}
  p_\pm= \frac{ m^2 +{\mathbf{p}}_\pm^2}{\tilde s y_\pm} \, \tilde p'
  + y_\pm  \tilde k_1 +  p_{\pm\perp}
  \,, \quad
  \tilde k_2= \frac{ \tilde{\mathbf{k}}_2^2}{\tilde s y_2} \, \tilde p' +
  y_2\tilde k_1 +\tilde k_{2\perp} \,.
  \label{sudcross}
\end{eqnarray}
Comparing the kinematic invariants $a_i$, $b_i$ with their
corresponding crossed invariants and  using the replacements
(\ref{replace}) and the Sudakov decomposition (\ref{sudcross})
we find the substitution rules for the energy fractions and
Euclidean 2--vectors (compare Ref.~\cite{KLS74})
\begin{eqnarray}
  \label{rep1}
  && x_1 \to \frac{1}{y_+} \,, \quad
     x_2 \to -\frac{y_2}{y_+} \,, \quad
     x_3 \to - \frac{y_-}{y_+} \,,
\\
  && \frac{{\mathbf{k}}_1}{x_1} \to  {\mathbf{p}}_+ \,, \quad
     \frac{{\mathbf{r}}_1}{x_1} \to -{\mathbf{p}}_- \,, \quad
     \frac{{\mathbf{k}}_2}{x_2} \to  {\mathbf{p}}_+ - \frac{y_+}{y_2}
  \tilde{\mathbf{k}}_2 \,, \quad
  \frac{{\mathbf{r}}_2}{x_2} \to -{\mathbf{p}}_- + \frac{y_-}{y_2}
  \tilde{\mathbf{k}}_2  \,,
\nonumber
\end{eqnarray}
with
\begin{eqnarray*}
{\mathbf{p}}_+ + {\mathbf{p}}_- + \tilde{\mathbf{k}}_2
+ {\mathbf{k}}  = 0 \,.
\end{eqnarray*}
To take into account the substitution rules for the polarizations
of the particles, we have to re-introduce the azimuthal angle of the
initial electron $\varphi_1$. This changes only the phase of the
factor $S_{\lambda_{e3}}$ given in Eq.~(\ref{S}).
Therefore, in this section we use
\begin{eqnarray}
\label{Scomplete}
  S_{\lambda_{e1}\lambda_{e3}}(\varphi_1,\varphi_3)&=&
  s \sqrt{x_3} \,{\mathrm{e}}^{{\mathrm{i}} \,
  (\varphi_3 \lambda_{e3} -\varphi_1 \lambda_{e1})}
  =4 E_1 E_2   \sqrt{x_3} \, {\mathrm{e}}^{{\mathrm{i}} \,
  (\varphi_3 \lambda_{e3} -\varphi_1 \lambda_{e1})}
  \,,
\\
  S_{\lambda_{e3}} &=& S_{\lambda_{e1}\lambda_{e3}}(0,\varphi_3)
\nonumber
\end{eqnarray}
instead of $S_{\lambda_{e3}}$.
Consequently, the helicity amplitudes given in Eqs.~(\ref{nonflip})
and (\ref{flip}) have to be  used now with the factor
$S_{\lambda_{e1}\lambda_{e3}}(\varphi_1,\varphi_3)$
and they depend on both lepton azimuthal angles explicitly.

Under crossing we have the substitution rules for the helicities
and the lepton azimuthal angles
\begin{eqnarray}
  \label{rep3}
  \lambda_1 \to - \lambda_1 \,, \quad
  \lambda_{e1} \to - \lambda_+ \,, \quad
  \lambda_{e3} \to   \lambda_- \,, \quad
  \lambda_2 \to \lambda_2 \,, \quad
  \varphi_{1,3} \to \varphi_{+,-} \,.
\end{eqnarray}
which leads to
\begin{eqnarray} \label{Schange}
  S_{\lambda_{e1}\lambda_{e3}}(\varphi_1,\varphi_3) \to
  \widetilde{S}_{\lambda_+ \lambda_-}& =&
  - {\mathrm{i}} \,  \tilde s \sqrt{y_+ y_-} \,
  {\mathrm{e}}^{{\mathrm{i}} \, (\varphi_- \lambda_- +\varphi_+
  \lambda_+)}\,.
\end{eqnarray}
The kinematic invariants are transformed to
\begin{eqnarray}
  \label{crosskin}
  \left\{ {a_1 \atop b_1} \right\}
  \to a_\pm&=&
  \pm 2 \tilde k_1 p_\pm=\pm \frac{1}{y_\pm}
  \left( m^2 +{\mathbf{p}}_\pm^2 \right) \,,
\nonumber
\\
  \left\{ {a_2 \atop b_2} \right\}
  \to b_\pm&=& \mp 2
  \tilde k_2 p_\pm= \mp \frac{1}{y_\pm y_2}
  \left[
      y_2^2m^2 +\left( y_2 {\mathbf{p}}_\pm -
       y_\pm \tilde {\mathbf{k}}_2 \right)^2
  \right] \,,
\\
  \left\{ {a \atop b} \right\}
  \to \tilde a_\pm&=&
  \mp k^2 \mp 2 k p_\mp=
  a_\pm + b_\pm \pm \frac{1}{y_2}\tilde{\mathbf{k}}_2^2 \,,
\nonumber
\\
  A \to y_- B &=&
  y_- \left\{ - \frac{1}{y_+ a_+ \ \tilde a_+}
  + \frac{1-y_+}{ y_+ a_+ \ y_- b_-} + \frac{1}{ y_- b_- \ \tilde a_-}
  \right\} \,.
\nonumber
\end{eqnarray}

We introduce the permutation operator ${\cal {P}}$
with the properties
\begin{eqnarray}
{\cal{P}} f(p_+,\lambda_+;p_-,\lambda_-)=
f(p_-,\lambda_-;p_+,\lambda_+)\,,
\end{eqnarray}
or in more detail
\begin{eqnarray}
{\cal{P}} f(a_\pm,b_\pm,\tilde a_\pm,y_\pm,\pi_\pm,\lambda_\pm)=
f( -a_\mp,-b_\mp,-\tilde a_\mp,y_\mp,\pi_\mp,\lambda_\mp)\,.
\end{eqnarray}
As usual, the circular components of ${\mathbf{p}}_\pm$ and $\tilde
{\mathbf{k}}_2$ are used
\begin{eqnarray}
  \pi_{\pm}= p_{x\pm} + {\mathrm{i}} \, p_{y\pm} \,, \quad
  \tilde\kappa_{2}=\tilde k_{2x} + {\mathrm{i}} \, \tilde k _{2y} \,.
\end{eqnarray}
The operator ${\cal{P}}$ is similar to ${\cal{P}}_{12}$ in the
original channel, in fact it will interchange
$e^+ \leftrightarrow e^-$ in the crossed amplitudes.

As a result, we obtain the amplitude for reaction
(\ref{crossreaction}) in the considered jet kinematics
\begin{eqnarray}
  M(\gamma e^\pm \to \gamma e^+e^- + e^\pm)=
  \frac{32 \pi^2 \alpha^2}{k^2}
  \widetilde{\cal{M}}_{\lambda_{+} \lambda_{-}}
  ^{\lambda_1  \; \lambda_2}
  {\mathrm{e}}^{-{\mathrm{i}} \, \lambda_{e4} \, \varphi_4}
  \,\delta_{ \lambda_{e2} ,\lambda_{e4}}\,,
  \label{fullampcross}
\end{eqnarray}
where $\lambda_{e2}$ ($\lambda_{e4}$) are the helicities of the
electrons or positrons moving along the $-z$ axis before and after
the collision with the initial photon. Using the function $C$ in
Eq.~(\ref{alternC}) for crossing and the notations
\begin{eqnarray}
  Y_1= B \pi_+ + \frac{\kappa}{y_- b_- \, \tilde a_-} \,, \quad
  Y_2= \pi_- - \frac{y_-}{y_2} \tilde \kappa_2\,,
\end{eqnarray}
the helicity amplitudes in the cross channel are presented as follows:
\begin{eqnarray}
  \label{crosshel}
  &&\widetilde{\cal{M}}_{-+}^{-+} =
    2 \widetilde{S}_{-+} \, y_+ (1 - {\cal{P}})
    \left\{ Y_1^*Y_2^*
    -\frac{ \pi_+^* \kappa^*}{y_+ a_+ \, \tilde a_+} \right\} \,,
\nonumber
\\
  &&\widetilde{\cal{M}}_{-+}^{+-} =
     \widetilde S_{-+}{\cal{P}}\left(\widetilde{\cal{M}}_{-+}^{-+}\over
     \widetilde S_{-+}\right)^*\,,
\nonumber
\\
  &&\widetilde{\cal{M}}_{-+}^{--} =
   -2 \widetilde{S}_{-+}
   \left( \widetilde{C} +
   {\cal{P}} \widetilde{D}^* \right) \,,
\nonumber
\\
  &&\widetilde {\cal{M}}_{-+}^{++} =
   \widetilde S_{-+}{\cal{P}}\left(\widetilde{\cal{M}}_{-+}^{--}
   \over\widetilde S_{-+}\right)^*\,,
\\
  &&\widetilde{D}=
  - (1-y_+) \left\{
      Y_1 Y_2^*  -\frac{ \pi_+ \kappa^*}{y_+ a_+ \, \tilde a_+}
  \right\} \,,
\nonumber
\\
  &&\widetilde{C}=
  -\frac{1-y_-}{1-y_+} \widetilde{D}^*
  -\frac{1}{\tilde a_+} + \frac{1}{\tilde a_-}
  +\frac{y_2  \left( m^2 - Y_2 \pi_+^* \right)}{y_+ a_+ \, y_- b_-}
  -\frac{\pi_+^* \tilde \kappa_2}{y_+ a_+ \, \tilde a_+}
  +\frac{\tilde \kappa_2^*Y_2}{y_- b_- \, \tilde a_-} \,,
\nonumber
\\
  &&\widetilde{\cal{M}}_{--}^{--} =
   2 m \widetilde{S}_{--} \,  (1+{\cal{P}})
  \left\{ B Y_2
     - \frac{ \kappa }{y_+ a_+ \, \tilde a_+} \right\} \,,
\nonumber
\\
  &&\widetilde{\cal{M}}_{--}^{++} =
  2 m \widetilde{S}_{--} \, y_2 \, (1+{\cal{P}}) Y_1 \,,
\nonumber
\\
  &&\widetilde{\cal{M}}_{--}^{-+} =
  2 m \widetilde{S}_{--} \, (1+{\cal{P}})
  \left\{
  (1-y_2) Y_1^* + B \frac{y_-}{y_2} \tilde \kappa_2^*
  + \frac{\kappa^* }{y_+ a_+ \, b_-}
  \right\} \,,
\nonumber
\\
  &&\widetilde{\cal{M}}_{--}^{+-} =  0 \,.
\nonumber
\end{eqnarray}
The remaining amplitudes are derived from a relation
equivalent to Eq.~(\ref{parity})
\begin{eqnarray}
  \widetilde{\cal{M}}_{-\lambda_+,-\lambda_-}^
  {-\lambda_1,\: -\lambda_2} =
  - (-1)^{\lambda_+ + \lambda_-}   \left(
  \widetilde{\cal{M}}_{\lambda_+,\lambda_-}^
  {\lambda_1, \: \, \lambda_2}\right)^*
  \,.
  \label{crossparity}
\end{eqnarray}
The helicity amplitudes (\ref{crosshel}), (\ref{crossparity}) are
$C$--odd. Therefore, they have to be symmetric under
lepton--antilepton exchange $e^- \leftrightarrow e^+$.
This symmetry, expressed by the relation
\begin{eqnarray}
  \widetilde{\cal{M}}_{\lambda_+ \lambda_-}^
  {\lambda_1 \: \lambda_2}
  \left(a_\pm,b_\pm,\tilde a_\pm,y_\pm,\pi_\pm,\varphi_\pm\right)&=& \\ \nonumber
&&\widetilde{\cal{M}}_{\lambda_- \lambda_+}^
  {\lambda_1 \: \lambda_2}
  \left(-a_\mp,-b_\mp,-\tilde a_\mp,y_\mp,\pi_\mp,\varphi_\mp\right)\,,
\end{eqnarray}
can be easily seen for all amplitudes taking
into account Eq.~(\ref{crossparity}) and ${\cal{P}}^2= 1$.

Note an alternative form for $\widetilde{C}$ useful for deriving
the hard photon limit discussed below in Section~\ref{crosslimits}:
\begin{eqnarray}
  \widetilde{C}=
  \frac{y_-}{1-y_+} \widetilde{D}^*
  - B \left( \frac{y_- \tilde \kappa_2 \pi_+^*}{ y_2} - m^2 \right)
  - \frac{\tilde \kappa_2 \kappa^*} {y_2 b_- \tilde a_-}
  - \frac{\kappa \pi_+^*}{y_+ a_+ b_-} \,.
  \label{Ccrossalt}
\end{eqnarray}

Since the function $B$ analogous to $A$ tends to zero
for $|{\mathbf{k}}|\to 0$,
the helicity amplitudes (\ref{crosshel}) are again proportional to
$|{\mathbf{k}}|$ in that limit as expected from crossing.
For $\widetilde{C}$ this can be seen using its form (\ref{Ccrossalt}).

%%%%%%%%%%%%%%%%%%%%%%%%%%%%%%%%%%%%%%%%%%%%%%%%%%%%%%%%%%%%%

\subsection{Soft and hard particle limits}
\label{crosslimits}

In the soft photon limit ($\tilde k_2\to 0$) we use
\begin{eqnarray}
  {\mathbf {Q}}=
  \frac{{\mathbf{p}}_+}{m^2+{\mathbf{p}}_+^2}+
  \frac{{\mathbf{p}}_-}{m^2+{\mathbf{p}}_-^2}
  \,, \quad
  R= \frac{1}{m^2+{\mathbf{p}}_+^2}-
  \frac{1}{m^2+{\mathbf{p}}_-^2}
  \label{Q+-R+-}
\end{eqnarray}
and have to consider (at $y_2\to 0$)
\begin{eqnarray}
  \tilde a_\pm \to a_\pm \,, \quad
  B \to \frac{R}{b_-} \,, \quad
  {\cal{P}} B
  \to \frac{R}{b_+} \,.
\end{eqnarray}
This limit again leads to a factorized form of the upper vertex:
\begin{eqnarray}
  J_{\mathrm{up}}(\gamma_{\lambda_1} \gamma^*
  \to e^+_{\lambda_{+}} e^-_{\lambda_{-}}\gamma_{\lambda_2})=
  \sqrt{4 \pi \alpha} \,
  J_{\mathrm{up}}(\gamma_{\lambda_1} \gamma^*
  \to e^+_{\lambda_{+}} e^-_{\lambda_{-}}   ) \, \widetilde{A}_2\,,
\end{eqnarray}
where
$J_{\mathrm{up}}(\gamma_{\lambda_1} \gamma^*
\to e^+_{\lambda_{+}} e^-_{\lambda_{-}})$ can be found in
Ref.~\cite{KSSS} and
\begin{eqnarray}
  \label{A2cross}
  \widetilde{A}_2=
  \left(\frac{p_+}{p_+ \tilde k_2}-\frac{p_-}{p_- \tilde k_2} \right)
  \!
  e_2^{*}
  = \frac{2}{y_2}
  \left( \frac{y_2 {\mathbf {p}}_+ - y_+
  \tilde {\mathbf {k}}_2 }{b_+}
   +\frac{y_2 {\mathbf {p}}_- - y_- \tilde {\mathbf {k}}_2 }{b_-}
  \right) \! {\mathbf{e}}^{(\lambda_2)*}
\end{eqnarray}
is the classical emission factor of the final photon.
With notations (\ref{Q+-R+-}) and (\ref{A2cross})
we find the non-vanishing helicity amplitudes in that limit
\begin{eqnarray}
  \label{softphotoncross}
  \widetilde{\cal{M}}_{\lambda_+ \lambda_-}^{\lambda_1 \: \lambda_2}
  &=& {\mathrm{i}} \, \tilde s \, \sqrt{y_+ y_-}
  {\mathrm{e}}^
  {{\mathrm{i}}\,(\varphi_+\lambda_+ +\varphi_-\lambda_-)}
  \widetilde{A}_2
\\
  &\times &
  \Bigg\{
  2 {\mathbf{Q}} {\mathbf{e}}^{(\lambda_1)} \,
  \left( y_+ - \delta_{\lambda_1, -2\lambda_+} \right)
  \delta_{\lambda_+,-\lambda_-}
  - \sqrt{2} m R
  \delta_{\lambda_1, 2\lambda_+} \delta_{\lambda_+,\lambda_-}
  \Bigg\} \,.
\nonumber
\end{eqnarray}

{}From Eq.~(\ref{softphotoncross}) we derive the amplitudes in
the limit of a hard electron ($y_-\to 1$, $y_+,y_2 \ll 1$)
\begin{eqnarray}
  \widetilde{\cal{M}}_{\lambda_+ \lambda_-}^{\lambda_1 \: \lambda_2}
  &=& -{\mathrm{i}} \,
  \tilde s \, \sqrt{y_+} {\mathrm{e}}^{ {\mathrm{i}} \, (\varphi_+
  \lambda_+ +\varphi_- \lambda_-)} \frac{2}{b_-} {\mathbf{p}}_-
  {\mathbf{e}}^{(\lambda_2)*}
\\
  &\times & \Bigg\{ 2
  {\mathbf{Q}} {\mathbf{e}}^{(\lambda_1)} \,
  \delta_{\lambda_+,-\lambda_-}
  + \sqrt{2} m R \delta_{\lambda_+,\lambda_-}
  \Bigg\} \delta_{\lambda_1, 2 \lambda_-} \,.
\nonumber
\end{eqnarray}
In this case the helicity of the initial photon is transferred to
that of the hard electron only, i.e., $\lambda_1=2 \lambda_-$.

The soft electron limit corresponds to the condition $y_-\ll 1$ and
finite transverse momentum ${\mathbf{p}}_-$. It can be obtained as
cross channel from Eq.~(\ref{softelectron})
performing the replacements (\ref{rep3}) and (\ref{Schange}) for
$S_{\lambda_{e1} \lambda_{e3}}$ as well as
\begin{eqnarray}
  &&x_1 \to \frac{1}{y_+} \,, \quad
  x_2 \to - \frac{y_2}{y_+}=-\frac{1-y_+}{y_+} \,,
\nonumber
\\
  &&{\mathbf{k}}_1 \to\frac{1}{y_+} {\mathbf{p}}_+ \,, \quad
  {\mathbf{k}}_2 \to - \frac{1-y_+}{y_+} {\mathbf{p}}_+ + \tilde
  {\mathbf{k}}_2 \,, \quad
  {\mathbf{p}}_3 \to {\mathbf{p}}_- \, ,
\\
  && a_1 \to a_+\,, \quad a_2 \to b_+\,, \quad
  a \to m^2 +( {\mathbf{p}}_+ + \tilde {\mathbf{k}}_2 )^2 \,, \quad
  x_3 b  \to m^2 + {\mathbf{p}}_-^2 \,.
\nonumber
\end{eqnarray}
In that limit the helicity amplitudes vanish as $\sqrt{y_-}$.

Finally we consider the hard final photon limit $y_2 \to 1$ and
$y_\pm \ll 1$. From the general helicity amplitudes
(\ref{crosshel}) and using the form (\ref{Ccrossalt}) of
$\widetilde{C}$ we find in leading order in $y_+$ and $y_-$
\begin{eqnarray}
  \widetilde{\cal{M}}_{-+}^{--}
  &=&
  \widetilde{\cal{M}}_{-+}^{++}
  =- 2 \widetilde{S}_{-+}
  \left\{ B \left( m^2 - \pi_-^* \pi_+\right)
  +\frac{\kappa^* \pi_+}{y_+ a_+ \, \tilde a_+}
  +\frac{\pi_-^* \kappa}{y_- a_- \, \tilde a_-} \right\}\,,
\nonumber
\\
  \widetilde{\cal{M}}_{--}^{++}
  &=&
  \widetilde{\cal{M}}_{--}^{--}
  =
  2 m \widetilde{S}_{--}
  (1+{\cal{P}}) \left\{ B \pi_- - \frac{\kappa}{y_+ a_+ \, \tilde a_+}
  \right\} \,,
\\
  \widetilde{\cal{M}}_{\lambda_+ \lambda_-}^{\,+\; -}
  &=&
  \widetilde{\cal{M}}_{\lambda_+ \lambda_-}^{\,-\; +}
  =0\,,
\nonumber
\end{eqnarray}
where in this limit $b_\pm \to - a_\pm$ and
\begin{eqnarray}
  B&=& {\cal{P}}B=- \frac{1}{y_+ a_+ \, \tilde a_+} -
     \frac{1}{y_+ a_+ \, y_- a_-} -
     \frac{1}{y_- a_- \, \tilde a_-}\,,
\\
  \tilde a_+ &=& - {\cal{P}} \tilde a_-= y_- a_+ + m^2 +
  ( {\mathbf{p}}_+ + \tilde {\mathbf{k}}_2)^2 \,,
\nonumber
\end{eqnarray}
the $a_\pm$ are defined in Eqs.~(\ref{crosskin}).
Note that the helicity of the initial photon is transferred to
that of the hard final photon, the amplitudes vanish as
$\sqrt{y_+ y_-}$.

%%%%%%%%%%%%%%%%%%%%%%%%%%%%%%%%%%%%%%%%%%%%%%%%%%%%%%%%%%%%%%%%%

\section{Conclusions}

The present work completes a series of
publications~\cite{KSS85,KSS86,KSSS} where the helicity analysis
of Born processes with non-decreasing cross sections up to
fourth order in $\alpha$ has been performed in a jet kinematics.
These specific kinematic conditions provide the main contribution
to the total cross section at high energies. On the other hand,
under these kinematic conditions Monte Carlo simulations as well as
analytical calculations by known methods
(compare, for example, Ref.~\cite{Behr})
are difficult to perform. Therefore, the results obtained can be
useful for monitoring and calibrating of polarized beams at high
energy colliders. They might also be used for almost exact estimates of
the essential background in a number of reactions with polarized
particles.

The main results of our paper are summarized in Eqs.~(\ref{fullamp}),
(\ref{nonflip}), (\ref{flip}) and (\ref{fullampcross}),
(\ref{crosshel}) which give the
analytical expressions for all 64 helicity amplitudes
to the high accuracy (\ref{accuracy}). Let us summarize
some specific features of those helicity amplitudes using
as example the double bremsstrahlung process:

\begin{itemize}
\item
The obtained formulae for the helicity amplitudes are compact and
convenient for numeric calculations since in their form large
compensating terms are already cancelled. Indeed, in
Eqs.~(\ref{nonflip}) and (\ref{flip}) all items are either
proportional to $\kappa = k_x +{\mathrm{i}} \, k_y$ or to the
function A (\ref{eq:16}) which vanishes at small transverse momentum
of the $t$--channel exchange photon $|{\mathbf {k}}|$. Therefore, the
amplitude
\begin{eqnarray}
  \vert M(e^-e^+\to e^-\gamma\gamma + e^+) \vert
  \propto \vert {\mathbf{k}} \vert \ \
  {\mathrm{ at}} \ \ \vert {\mathbf {k}} \vert \to 0 \,.
\end{eqnarray}

\item
The helicity amplitudes ${\cal M}_{+\;-}^{\lambda_1 \lambda_2}$ and
${\cal M}_{-\;+}^{\lambda_1 \lambda_2}$, in which the electron changes
its helicity, are proportional to the electron mass $m$
(see Eq.~(\ref{flip})). They become small compared with the non-spin
flip amplitudes ${\cal M}_{+\;+}^{\lambda_1 \lambda_2}$ and
${\cal M}_{-\;-}^{\lambda_1\lambda_2}$  for large scattering angles
(\ref{langle}) or small energies of both photons
$\omega_{1,2}\ll E_1$.

\item
The amplitudes ${\cal M}_{+-}^{--}$ and ${\cal M}_{-+}^{++}$ with the
maximal change of helicities
$|\lambda_{e1}-\lambda_{e3}-\lambda_{1}-\lambda_{2}|=3$
are equal to zero under our kinematic conditions.

\item
If the energy of any final particle tends to that of the initial
electron, the incoming electron transfers its helicity
to that particle: $\lambda_i = 2 \lambda_{e1}$ at $\omega_i
\to E_1$ or $\lambda_{e3} = \lambda_{e1}$ at $E_3 \to E_1$.
\end{itemize}

%%%%%%%%%%%%%%%%%%%%%%%%%%%%%%%%%%%%%%%%%%%%%%%%%%%%%%%%%%

\section*{ Acknowledgements}

This work is supported in part by Volkswagen Stiftung (Az. No.
I/72 302) and by Russian Foundation for Basic Research (codes
96-15-96030, 99-02-17211 and 99-02-17730).

%%%%%%%%%%%%%%%%%%%%%%%%%%%%%%%%%%%%%%%%%%%%%%%%%%%%%%%%%%

\appendix
\section{Calculation of necessary bilinear spinor structures}
\label{appa}
\setcounter{equation}{0}
\renewcommand{\theequation}{A.\arabic{equation}}

We have to  calculate all bilinear spinor structures appearing in
Eq.~(\ref{eq:18}) in order to find the helicity amplitudes in
the $s\to \infty$ limit.
For example, let us consider the numerator of the second
term of Eq.~(\ref{eq:18}).
\begin{eqnarray}
  N=\bar u_3  \hat p' \hat k_{\perp} \hat e_2^*
  (\hat p_1-\hat k_1+m) \hat e_1^* u_1 \,.
\end{eqnarray}
We transpose $\hat e_1^*$ to the left and obtain
\begin{eqnarray}
  N=\bar u_3  \hat p' \hat k_{\perp} \hat e_2^*
  \left[ \hat e_1^* (-\hat p_1+\hat k_1 + m) + 2 p_1 e_1^* \right]
  u_1 \,.
\end{eqnarray}
Using the Sudakov decomposition of the 4--vectors $e_i^*$ and $k_1$
[see Eqs.~(\ref{sudakov}), (\ref{polvector}), (\ref{eq:10})]
and taking into account $\hat p' \hat p'=0$,
$ \hat p u_1 \approx \hat p_1 u_1 = m u_1$
we find
\begin{eqnarray}
  N=\bar u_3  \hat p' \hat k_{\perp} \hat e_{2\perp}^*
  \left[ \hat e_{1\perp}^* \hat k_{1\perp}
   + m x_1 \hat e_{1\perp}^*  +
  \frac{2}{x_1} {\mathbf{k}}_1 {\mathbf{e}}^{(\lambda_1)*}
  \right] u_1 \,.
\end{eqnarray}
Similar transformations can be performed for the other terms of
Eq.~(\ref{eq:18}).

As a result we conclude that it is sufficient to calculate the
following  combinations
$\bar u_3 \hat p' \hat a_\perp \hat b_\perp \dots \hat c_\perp u_1$
and $\bar u_3 \hat a_\perp \hat b_\perp \dots \hat c_\perp u_1$.
Here the bispinors $u$ are defined as follows (for an initial
electron with helicity +1/2)
\begin{eqnarray}
  u_1=\left( \begin{array}{c}
           \sqrt{E_1 +m }\, w_1\\
           \sqrt{E_1 -m }\, w_1
           \end{array} \right)
  \,, \ \
  u_3=\left( \begin{array}{c}
           \sqrt{E_3 +m }\, w_3\\
           2 \lambda_{e3}\sqrt{E_3 -m } \,w_3
           \end{array} \right)\,,
\end{eqnarray}
with
\begin{eqnarray}
  w_1=\left(\begin{array}{c}1\\
                          0  \end{array}\right)
  \,, \ \
  w_3^{(+1/2)}=\left(\begin{array}{c}{\mathrm{e}}^{-{\mathrm{i}} \,
  \varphi_3/2}
\\
  \frac{\theta_3}{2}\,
  {\mathrm{e}}^{{\mathrm{i}}\varphi_3/2}\end{array}\right)
  \,, \ \
  w_3^{(-1/2)}=\left(\begin{array}{c}-\frac{\theta_3}{2}\,
  {\mathrm{e}}^{-{\mathrm{i}}\varphi_3/2}\\
  {\mathrm{e}}^{{\mathrm{i}}\varphi_3/2}\end{array}\right)
  \,,
\nonumber
\\
\end{eqnarray}
and
\begin{eqnarray}
  \hat p'=E_2
  \left(\begin{array}{cc}1&\sigma_z\\-\sigma_z&-1\end{array}\right)
  \,, \ \
  \hat a_{\perp}= - a_x \gamma_x - a_y \gamma_y \,.
\end{eqnarray}
The $\sigma_i$ and $\gamma_i$ are the Pauli and Dirac matrices in the
standard representation. Introducing the circular components of the
Euclidean 2--vector $\mathbf{a}$
\begin{eqnarray}
  a_\pm= a_x \pm {\mathrm{i}} \, a_y\,, \qquad a_\mp^*=a_\pm\,,
\end{eqnarray}
we obtain
\begin{eqnarray}
  \hat a_\perp= - \frac{1}{2} (a_+ \gamma_- + a_- \gamma_+) \,, \quad
  \gamma_\pm=\left( \begin{array}{cc}
            0 &\sigma_\pm\\
            -\sigma_\pm & 0
           \end{array} \right) \,.
\end{eqnarray}

Since $\sigma_+ w_1=0$ and $\sigma_- w_1= { 0 \choose 2}$, we have
$ \gamma_+ u_1= 0$  and $\gamma_+ \gamma_- u_1= -4 u_1$. From this it
follows that
\begin{eqnarray}
  &&\hat a_\perp u_1= - \frac{1}{2} a_+ \gamma_- u_1\,, \quad
  \hat a_\perp \hat b_\perp u_1= - a_- b_+ u_1\,,
\\
  &&\hat a_\perp \hat b_\perp \hat c_\perp u_1= \frac{1}{2}
  a_+ b_- c_+ \gamma_- u_1 \,, \ \ \
  \hat a_\perp \hat b_\perp \hat c_\perp \hat d_\perp u_1=
   a_- b_+ c_-  d_+   u_1 \,.
\nonumber
\end{eqnarray}
Introducing the shorthand notations
\begin{eqnarray*}
  \delta_{\lambda_{e3},\pm 1/2}=\delta_{\pm} \,,\quad
  \delta_{\lambda_i,\pm 1}=\delta_{i\pm}
\end{eqnarray*}
and using the quantity $S_{\lambda_{e3}}$ defined in Eq.~(\ref{S})
we arrive at the following four basic combinations
\begin{eqnarray}
  \bar u_3 \hat p'  u_1 &=& S_+ \delta_+ \,,  \quad
  \bar u_3 \hat p' \hat a_\perp u_1= - a_+ S_- \delta_- \,, \quad
\nonumber
\\
  s \,\bar u_3 u_1 &=&
  \frac{1}{x_3} \left[
  - \pi_3 S_- \delta_- + (1+x_3) m S_+ \delta_+ \right]
  \,,
\\
  s \, \bar u_3 \hat a_\perp u_1 &=&
   - \frac{a_+}{x_3} \left[
  \pi_3^* S_+ \delta_+ + (1-x_3) m S_- \delta_- \right]\,,
\nonumber
\end{eqnarray}
where
\begin{eqnarray}
  \pi_3=p_{3x} +{\mathrm{i}} \, p_{3y}
       =-\kappa -\kappa_1 - \kappa_2 \,.
\end{eqnarray}
Choosing $a_\perp= e_\perp^{(\lambda) *}$ we find
$a_\pm= \mp \sqrt{2} \delta_{\lambda,\pm 1}$.

Then the relevant bilinear combinations  [using the Sudakov
decomposition of all 4--vectors of Eq.~(\ref{eq:18})] are found to be
\begin{eqnarray}
  \bar u_3\hat p'u_1&=& S_+ \delta_+ \,,
\nonumber
\\
  \bar u_3 \hat p'\hat e_i^* u_1&=& \sqrt{2}\delta_{i+}S_-\delta_- \,,
\nonumber
\\
  \bar u_3 \hat p' \hat e_2^* \hat e_1^*   u_1&=&
  2  \delta_{2-}\delta_{1+} S_+\delta_+ \,,
\nonumber
\\
  \bar u_3 \hat p' \hat e_{2\perp}^* \hat k_{i\perp}   u_1&=&
  - \sqrt{2} \kappa_i \delta_{2-} S_+\delta_+  \,,
\nonumber
\\
  \bar u_3 \hat p' \hat k_{\perp} \hat e_{i\perp}^* u_1&=&
  \sqrt{2}  \kappa^* \delta_{i+} S_+\delta_+  \,,
\nonumber
\\
  \bar u_3 \hat p' \hat k_{\perp} \hat e_{2\perp}^*
  \hat e_{1\perp}^* u_1&=&
  -2 \kappa \delta_{2-} \delta_{1+} S_-\delta_- \,,
\\
  \bar u_3 \hat p' \hat e_{2\perp}^* \hat e_{1\perp}^*
  \hat k_{i\perp} u_1&=&
  -2 \kappa_i \delta_{2+} \delta_{1-} S_-\delta_-  \,,
\nonumber
\\
  \bar u_3 \hat p' \hat e_{2\perp}^* \hat k_{\perp}
  \hat e_{1\perp}^* u_1&=&
  2 \kappa^* \delta_{2+} \delta_{1+} S_-\delta_- \,,
\nonumber
\\
  \bar u_3 \hat p' \hat e_{2\perp}^* \hat e_{1\perp}^*
  \hat k_{\perp} u_1&=&
  -2 \kappa \delta_{2+} \delta_{1-} S_-\delta_-   \,,
\nonumber
\\
  \bar u_3\hat p'\hat e_{2\perp}^* \hat k_{\perp}
  \hat e_{1\perp}^* \hat k_{i\perp} u_1&=&
  2  \kappa \kappa_i \delta_{2-} \delta_{1-} S_+\delta_+  \,,
\nonumber
\\
  \bar u_3 \hat p' \hat k_{i\perp}\hat e_{2\perp}^*
  \hat e_{1\perp}^* \hat k_{\perp} u_1&=&
  -2 \kappa_i^* \kappa \delta_{2+} \delta_{1-} S_+\delta_+  \,,
\nonumber
\\
%%%%
  s \,\bar u_3 \hat e_{i\perp}^* u_1&=&
  \sqrt{2} \delta_{i+} \frac{1}{x_3}
  \left[ \pi_3^* S_+ \delta_+ + (1-x_3) m S_- \delta_-\right] \,,
\nonumber
\\
  s \,\bar u_3 \hat e_{2\perp}^* \hat e_{1\perp}^* u_1&=&
  2 \delta_{2-} \delta_{1+} \frac{1}{x_3}
  \left[ - \pi_3 S_- \delta_- + (1+x_3) m S_+ \delta_+\right] \,,
\nonumber
\\
  s \, \bar u_3 \hat e_{2\perp}^* \hat e_{1\perp}^*
  \hat k_{i\perp} u_1&=&
  -2 \delta_{2+} \delta_{1-} \frac{\kappa_i}{x_3}
  \left[ \pi_3^* S_+ \delta_+ + (1-x_3) m S_-  \delta_-\right]  \,,
\nonumber
\\
  s \,\bar u_3 \hat e_{2\perp}^* \hat k_{\perp}
  \hat e_{1\perp}^* u_1&=&
  2 \delta_{2+} \delta_{1+} \frac{\kappa^*}{x_3}
  \left[ \pi_3^* S_+ \delta_+ +(1-x_3) m S_- \delta_- \right]  \,.
\nonumber
\end{eqnarray}

%%%%%%%%%%%%%%%%%%%%%%%%%%%%%%%%%%%%%%%%%%%%%%%%%%%%%%%%%%%%%%%%%%%%

\end{document}